\documentclass[epj]{svjour}

\usepackage[british]{babel}
\usepackage{epsfig}
\usepackage{appendix}
\usepackage[switch,columnwise]{lineno}

\usepackage[space]{cite}
\usepackage{amsmath}
\usepackage{multirow}

\newcommand\ba{\begin{eqnarray}}
\newcommand\ea{\end{eqnarray}}
\newcommand\nn{\nonumber}
\newcommand{\be}{\begin{equation}}
\newcommand{\ee}{\end{equation}}

\usepackage{xspace}
\usepackage{graphics}
\graphicspath{{./images/}}
\usepackage{float}
\usepackage{adjustbox}
\usepackage{hyperref}
\usepackage{listings} 
\usepackage{color}
\definecolor{green}{rgb}{0,0.5,0}
\definecolor{blue}{rgb}{0,0,1}
\definecolor{red}{rgb}{1,0,0}
\definecolor{magenta}{rgb}{0.78,0.09,0.98}
\usepackage[T1]{fontenc}
\usepackage{lmodern}
\usepackage[utf8]{inputenc}  
\usepackage{subcaption}
\usepackage{authblk}

\usepackage{eurosym}

\newcommand{\PANDA}{$\overline{\textrm{P}}\textrm{ANDA}$\xspace}

\begin{document}\sloppy
\raggedbottom

\hugehead
%

\title{Feasibility studies for the measurement of time-like proton electromagnetic form factors from $\bar{p}p \rightarrow \mu^+\mu^-$ at \PANDA at FAIR}

%
%
\author{
G.~Barucca\inst{1} \and 
F.~Davì\inst{1} \and 
G.~Lancioni\inst{1} \and 
P.~Mengucci\inst{1} \and 
L.~Montalto\inst{1} \and 
P. P.~Natali\inst{1} \and 
N.~Paone\inst{1} \and 
D.~Rinaldi\inst{1} \and 
L.~Scalise\inst{1} \and 
W.~Erni\inst{2}
% ...
}
\institute{
Università Politecnica delle Marche-Ancona,{ \bf Ancona}, Italy \and 
Universität Basel,{ \bf Basel}, Switzerland
% ...
}

\date{Received: date / Revised version: date}
%
\abstract{
This paper reports on Monte Carlo simulation results for future measurements of time-like proton electromagnetic form factors, $|G_{E}|$ and $|G_{M}|$, using the $\bar{p} p \rightarrow \mu^{+} \mu^{-}$ reaction at \PANDA (FAIR). The electromagnetic form factors are fundamental quantities parameterizing the electric and magnetic structure of hadrons. This work estimates the statistical and total accuracy with which the form factors can be measured at \PANDA, using an analysis of simulated data within the PandaRoot software framework. The most crucial background channel is $\bar{p} p \rightarrow \pi^{+} \pi^{-}$, due to the very similar behavior of muons and pions in the detector. The suppression factors are evaluated for this and all other relevant background channels at different values of antiproton beam momentum. The signal/background separation is based on a multivariate analysis, using the Boosted Decision Trees method. An expected background subtraction is included in this study, based on realistic angular distributions of the background contribution. Systematic uncertainties are considered and the relative total uncertainties of the form factor measurements are presented.
}

\maketitle

\section{Introduction} \label{sec:intro}

  Electromagnetic form factors (FFs) are fundamental quantities which describe the internal structure of hadrons. The proton structure at leading order in $\alpha$ ($\alpha$ being the electromagnetic fine structure constant) can be described by the electric ($G_E$) and the magnetic ($G_M$) FFs. Experimental access to these FFs is possible via the measurement of differential and total cross sections for elastic electron-proton scattering in the space-like region (momentum transfer squared $q^2 < 0$~(GeV/$c$)$^2$), while in the time-like region ($q^2>0$~(GeV/$c$)$^2$), proton FFs can be accessed in annihilation processes of the type $\bar{p} p \rightarrow \ell^{+} \ell^{-}$ with $\ell$=$e,\mu, \tau$ (or the time-reversed process in case of electrons). Here the interaction takes place through the exchange of a single virtual photon, carrying a momentum transfer squared $q^2$.
 
 Although the space-like FFs have been studied since the 1950's \cite{Hof:1956}, the recent application of the polarization transfer method \cite{Ak:1968, Akhiezer:1974em} triggered new efforts in the field of electromagnetic proton FFs. Precise data on polarised elastic electron-proton scattering up to $Q^2$=$-q^2 \approx 8.5$~(GeV/$c$)$^2$ \cite{Jo:2000,Ga:2002,Pun:2005, Pu:2010,Puckett:2017flj} are in tension with the existing results obtained with the well-established Rosenbluth method \cite{Rosenbluth:1950yq}. The polarization transfer method showed that the ratio $\mu_p  G_E$/$G_M$ (where $\mu_p$ stands for the proton magnetic moment) decreases linearly from unity to zero with increasing values of $Q^2$.

At low momentum transfer, space-like FFs provide information on the distributions of the electric charges and magnetization within the proton. The proton charge radius is related to the derivative of the electric FF at $Q^2=0$~(GeV/$c$)$^2$. It has been determined from electron-proton scattering measurements and hydrogen spectroscopy \cite{Mohr:2008fa,Pohl:2010,Bezginov:2019mdi} but the results are not totally in agreement. Future experiments, e.g. the MUon proton Scattering Experiment (MUSE) at Paul Scherrer Institute (PSI) \cite{Gilman:2017hdr} aim to extend the current studies by determining the proton radius using both muon and electron scattering measurements.
In the time-like region, the proton FFs have been measured in electron-positron annihilation $e^+e^-\to \bar p p$ and proton-antiproton annihilation $\bar p p\to e^+e^- $~\protect\cite{Castellano:1973,Andreotti:2003,Ambrogiani:1999,Antonelli:1998fv,Bardin:1994am,Armstrong:1993,Delcourt:1979,Bisello:1983,Bisello:1990,Ablikim:2005,Ablikim:2015,Pedlar:2005,Akhmetshin:2015ifg}. In addition, the radiative return process $e^+e^-\to \bar p p \gamma$, where $\gamma$ is a hard photon emitted by initial state radiation (ISR), has been used by the BaBar and the BESIII collaboration to measure the time-like proton FF ratio $|G_E|/|G_M|$ and the effective FF $|F_p|$ in a continuous range of $q^2$ \protect\cite{Lees:2013b,Lees:2013,Ablikim:2019njl}. The data show some regular oscillations in the measured $|F_p|$, which are currently the subject of several theoretical studies. The precision of the measurements of the proton FFs $|G_E|$ and $|G_M|$ (and their ratio) in the time-like region has been limited over the past decades by poor statistics, in contrast to the space-like region measurements. In 2019 the BESIII collaboration measured the Born cross section of the $e^+e^-\to \bar p p$ process and the proton FFs at 22 center-of-mass (CM) energy points from $q^2=4$ (GeV/$c$)$^2$ to $q^2=9.5$ (GeV/$c$)$^2$ \cite{Ablikim:2019eau}. The FF ratio was determined with total uncertainties around $10\%$, comparable to the data in the space-like region at the same $|q^2|$ values.

Proton FFs in the space-like and time-like regions are connected via dispersion relations. Therefore, a precise determination of the time-like FFs over a large $q^2$ range using different electromagnetic processes can help to constrain the theoretical models and shed light on the discrepancies which have been found in the space-like region. The situation in the time-like region will be  improved even more in the future by the data which will be collected with the \PANDA (antiProton ANnihilation at DArmstadt)  detector.

The time-like proton FFs will be measured at \PANDA in the $\bar p p \to e^+ e^-$ and $\bar p p \to \mu^+ \mu^-$   annihilation processes  \cite{TomasiGustafsson:2008gq,Dbeyssi:2011tv}. It will be the first time that muons in the final state will be used to measure the time-like FFs of the proton. In contrast to the $\bar{p}p \rightarrow e^+e^-$ process, the  $\bar p p \to \mu^+ \mu^-$ reaction has the advantage that corrections due to final state radiation are expected to be smaller. Measuring both channels should therefore allow the formalism for radiative corrections to be tested. Moreover, a test of lepton universality at a few percent level could be possible at \PANDA, based on the determination of the effective FF of the proton with both channels.

The possibility to access the proton FFs in the region below the kinematic threshold of the proton antiproton production through the measurement of  the $\bar p p \to \ell^+ \ell^- \pi^0$ process \cite{Dubnickova:1995ns,Adamuscin:2006bk,Guttmann:2013} is under investigation. This region below ($2M_p$)$^2$ is called the unphysical region and it has never been experimentally accessed. Feasibility studies of exploiting the $\bar p p \to e^+ e^-$ reaction at \PANDA were addressed in Refs.~\cite{Sudol:2010,DmitryAlaaPaper}. It has been shown that a separate measurement of $|G_E|$ and  $|G_M|$ can be performed up to $q^2 \sim 14$ (GeV/$c$)$^2$.  In this paper, the results of a feasibility study to extract the time-like proton FFs using the  $\bar p p \to \mu^+ \mu^-$ process at \PANDA are presented.

\section{The \PANDA experiment at FAIR}
 \label{sec:panda}

The \PANDA experiment \cite{Lutz:2009ff} will be located at the Facility for Antiproton and Ion Research (FAIR), which is currently under construction in Darmstadt (Germany). The \PANDA experiment will measure annihilation reactions induced by a high-intensity antiproton beam covering a wide range of momenta between 1.5 GeV/$c$ and 15 GeV/$c$. The physics program includes hadron spectroscopy in the charmonium, hyperon and light quark sectors, hypernuclear physics, and studies of hadron properties in a nuclear medium. An important part of the \PANDA physics program will be dedicated to the investigation of the nucleon structure. It is planned to measure nucleon-to-meson transition distribution amplitudes (TDAs)  through the measurements of the exclusive processes $\bar p p \to \gamma^* \pi^0 \to e^+ e^- \pi^0$  \cite{Singh:2015} and $\bar p p \to J/\Psi \pi^0 \to e^+ e^- \pi^0$  \cite{Singh:2016qjg}.  The generalized distribution amplitudes (GDAs) of the proton can be also accessed with the large angle production of the neutral states $\gamma \gamma$ and $\pi^0 \gamma$ \cite{Lutz:2009ff}. In addition, a Drell-Yan physics program to access transverse momentum dependent (TMD) parton distribution functions (PDFs), using the inclusive production of lepton pairs in proton-antiproton annihilations, is also foreseen \cite{Lutz:2009ff}.

\begin{figure*}[ht]
\begin{center}
\resizebox{0.85\textwidth}{!}{
 \includegraphics{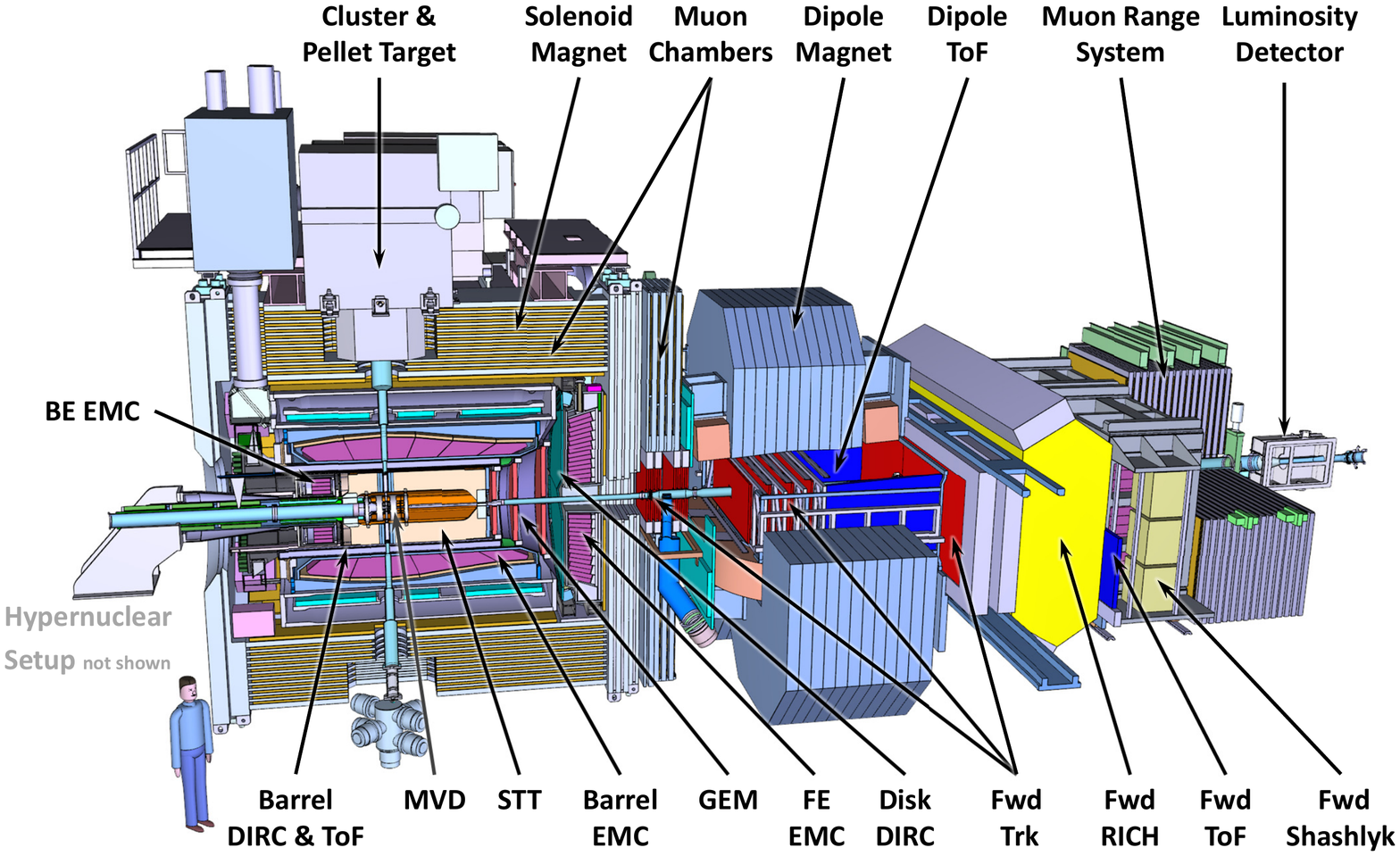}}
\caption{The proposed \PANDA detector.}
\label{fig:panda}
\end{center}
\end{figure*}

\subsection{The FAIR accelerator complex}

The FAIR accelerator complex will extend the existing facilities of the GSI Helmholtzzentrum  f\"ur  Schwerionenforschung in Darmstadt \cite{FAIR:2006}. It will deliver antiproton and ion beams for four main experimental projects that aim to provide fundamental insights into the structure of matter and the evolution of the universe. FAIR will use a new large  synchrotron ring, named SIS100, with a circumference of 1.1 km and a bending power of 100 Tm to accelerate protons up to 30 GeV/$c$. The accelerated protons will hit a copper target to produce antiproton beams with a time-averaged production rate in the $5.6\times 10^6$ to $10^7$ per second range. The antiprotons will be collected and cooled in the collector ring (CR), followed by their accumulation in the recycled experimental storage ring (RESR). Finally, the antiprotons will be injected in the high energy storage ring (HESR) where the induced antiproton annihilation reactions will be studied by the \PANDA fixed-target experiment.

This setup is designed to provide a beam of up to $10^{11}$ antiprotons per filling and peak instantaneous luminosities  up to $2 \times 10^{32}$ cm$^{-2}$ s$^{-1}$. However, in the initial start-up phase of the FAIR operation without the RESR, the HESR will be used as an accumulator, resulting in a luminosity about a factor of 20 lower than the nominal design value. In the present analysis, the results are obtained assuming an integrated luminosity of 2~fb$^{-1}$, which can be accumulated in four to five months of data taking at the maximum design luminosity.

\subsection{The \PANDA detector}

The proposed \PANDA detector \cite{Lutz:2009ff},  shown in Fig.~\ref{fig:panda}, will be located at the HESR. It is divided into a target spectrometer surrounding the target area and a forward spectrometer designed to detect particles in the forward rapidity region. The two spectrometers have  a solid angle acceptance of almost $4\pi$.

 The antiproton beams at the HESR will interact with a fixed proton target at CM energies between 2.2  and 5.5 GeV. A frozen pellet  and a cluster-jet  are two alternative hydrogen targets foreseen for the \PANDA $\bar p p$ annihilation studies \cite{PandaTDRTarget:2014}. In addition, internal targets filled by heavier gases and non-gaseous nuclear targets will be available for the $\bar p A$ studies and hypernuclear experiments, respectively.

 The target spectrometer is equipped with a superconducting solenoid magnet with a maximum magnetic field of 2 T \cite{PandaTDRMagnet:2009}. The innermost tracking system of the target spectrometer is the Micro Vertex Detector (MVD) \cite{PandaTDRMVD:2012}. It is based on radiation-hard silicon pixel and silicon strip sensors and is optimized for the detection of secondary decay vertices of short lived particles such as D-mesons and hyperons. The MVD will provide precise vertex position measurements with a resolution of about 100 $\mu$m  along  the beam axis and 30 $\mu$m  in the perpendicular plane. The Straw Tube Tracker (STT) is the central tracking detector in the target spectrometer \cite{PandaTDRStt:2013}. It encloses the MVD and is followed by three planar stations of Gas Electron Multipliers (GEM) downstream of the target. The MVD, the STT and the GEM will provide momentum measurement of charged particles with a transverse momentum resolution better than $1\%$. In addition, the measurement of the energy  loss by the STT and the MVD will be used for particle identification. A barrel and an end-cap Detection of Internally Reflected Cherenkov light (DIRC) detectors will be used  to separate pions from kaons at polar angles  between $5^{\circ}$ and $140^{\circ}$, and momenta up to 4 GeV/$c$ \cite{PandaDIRC:2005}. A time-of-flight (TOF) system, made of small plastic scintillator tiles (SciTil), will be also employed for particle identification of pions, protons and kaons.
 
 The energy of photons and electrons will be measured by an electromagnetic calorimeter (EMC), consisting of lead tungstate (PbWO$_4$) crystals operated at a temperature of $-25^{\circ}$ C to improve the light yield \cite{PandaTDREMC:2008}. Muon PID will be provided by the Muon System (MS), surrounding the solenoid magnet \cite{PandaTDRMDT:2012}. For the separation of muons from other particles, the range measurement technique is used, which is based on a sampling structure of active and passive layers in all subsystems of the MS. The MS is the most relevant component for this analysis and is described in detail in Ref. \cite{PandaTDRMDT:2012} .

 The forward spectrometer \cite{Lutz:2009ff,PandaTDRMagnet:2009,PandaTDRFSC:2016} with a 2 Tm dipole magnet will detect particles with polar angles below the end cap coverage of the target spectrometer. It comprises a forward tracking system (FTS), an Aerogel Ring Imaging Cherenkov Counter (FRICH), a Forward TOF system (FTOF), a Shashlyk calorimeter and a Forward Range System (FRS).  The forward spectrometer is completed by  the Luminosity Monitor Detector (LMD)  for precise determination of the absolute and the relative time integrated luminosities. A detailed overview of the \PANDA detector can be found in Ref.~\cite{Lutz:2009ff}.

In order to reach the physics goals of the experiment, operation at high event rates exceeding 20 $\cdot$ 10$^{6}$ $s^{-1}$ is expected. This requires a novel approach to data acquisition and real-time event selection. After a full online reconstruction of the events, an event filtering procedure based on a preliminary selection of the physics channels of interest will be performed.

\section{Reaction kinematics and cross sections} 
\label{sec:kine}

\begin{figure}[h]
    \centering
\resizebox{0.4\textwidth}{!}{
  \includegraphics{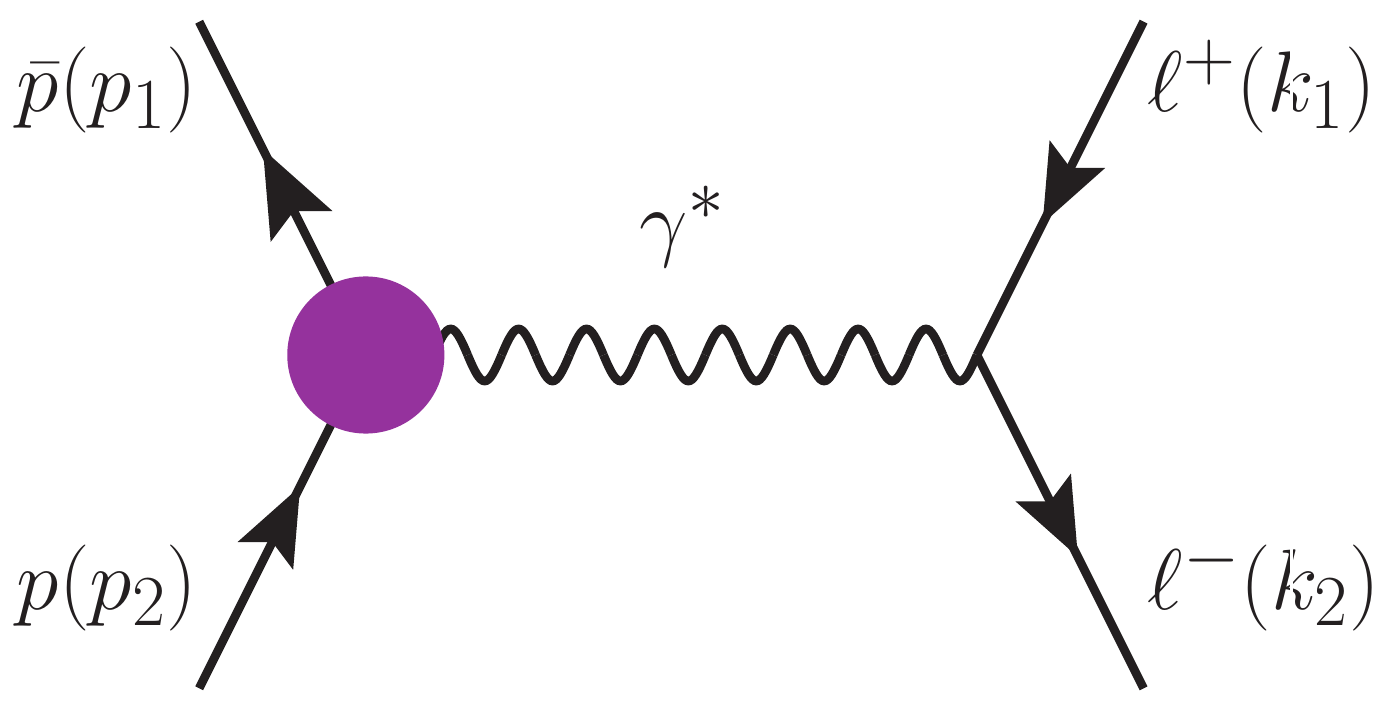}
}
\caption{Lowest-order QED diagram contributing to the reaction amplitude of antiproton-proton annihilation into $\ell^{+}\ell^{-}$ final states.}
\label{fig:feynman_tl}     
\end{figure}

The lowest-order QED contribution to the amplitude of the  $\bar{p}p \rightarrow {\ell}^+{\ell}^-$ $({\ell} = e, \mu)$  annihilation reaction is shown in Fig. \ref{fig:feynman_tl}. The four momenta of the involved particles are written in parentheses. Four-momentum conservation at the hadronic vertex implies that $q^2$ is equal to the $\bar{p}p$ CM energy squared $s$:
\begin{equation}
q^2 = (p_1 + p_2)^2 = s.
\end{equation}

In the Born approximation, which assumes one photon exchange, the differential cross section in the $\bar{p}{p}$ CM system of the annihilation of $\bar{p}p$ into a lepton pair can be written as a function of the Sachs FFs \cite{Zichichi:1962ni, Dbeyssi:2011tv}  as
\begin{equation}
\begin{aligned}
\frac{d\sigma}{d\cos {\theta_{CM}}} =& \frac{\pi {\alpha}^2}{2s} \frac{\beta_{\ell}}{\beta_{p}} 
\bigg[ \frac{1}{\tau} \bigg(1- {{\beta}_{\ell}}^2 \cos^{2}\theta_{CM}\bigg) {|G_E|}^2 \\ + & \bigg(2 - {\beta}_{\ell}^2 + {{\beta}_{\ell}}^2 \cos^{2}\theta_{CM}\bigg) {|G_M|}^2\bigg],
\label{eq:B6}
\end{aligned}
\end{equation}
where $\theta_{CM}$ is the polar angle of the  negative charged lepton $\ell^-$ and is measured with respect to the antiproton direction in the $\bar{p}p$ CM frame. $\alpha \approx$ 1/137 is the fine structure constant and the kinematic factors are
\begin{align*}
\beta_{\ell,p} &= \sqrt{1-4M^2_{\ell,p}/s},\\
\tau &= \frac{q^2}{4M_p^2},
\end{align*}
where $\beta_{\ell,p}$ is the velocity of the lepton or the proton in the CM frame, respectively. The measurement of the angular distribution of the charged leptons at a fixed energy requires a high luminosity in order to collect enough statistics over the whole angular range. With the precise knowledge of the luminosity, the absolute value of the cross section can be determined and an individual extraction of the time-like electromagnetic proton FFs, $|G_E|$ and $|G_M|$, is possible. 

The effective proton FF is a quantity which can be determined even at low statistics experiments. It is a linear combination of the $|G_E|$ and $|G_M|$ FFs, and can be obtained by the measurement of the integrated cross section ($\sigma(q^2)$) via
\begin{equation}
\label{eq:sigma_effFF}
\sigma(q^2) = \frac{4\pi{\alpha}^{2}}{q^2}  \bigg(1 -  \frac{ {\beta}_{\ell}^{2}}{3} \bigg ) \; \frac{\beta_{\ell}}{\beta_{p}} \bigg(1 +  \frac{1}{2\tau} \bigg ) |F_{p}|^2,
\end{equation}
being
\begin{equation}
\label{eq:effFF}
|F_{p}|  = \sqrt{\frac{2\tau {|G_{M}|^{2}}+|G_{E}|^{2}}{2\tau + 1}}.
\end{equation}

\begin{figure}[h]
\includegraphics[height=7.8cm,width=9.5cm]{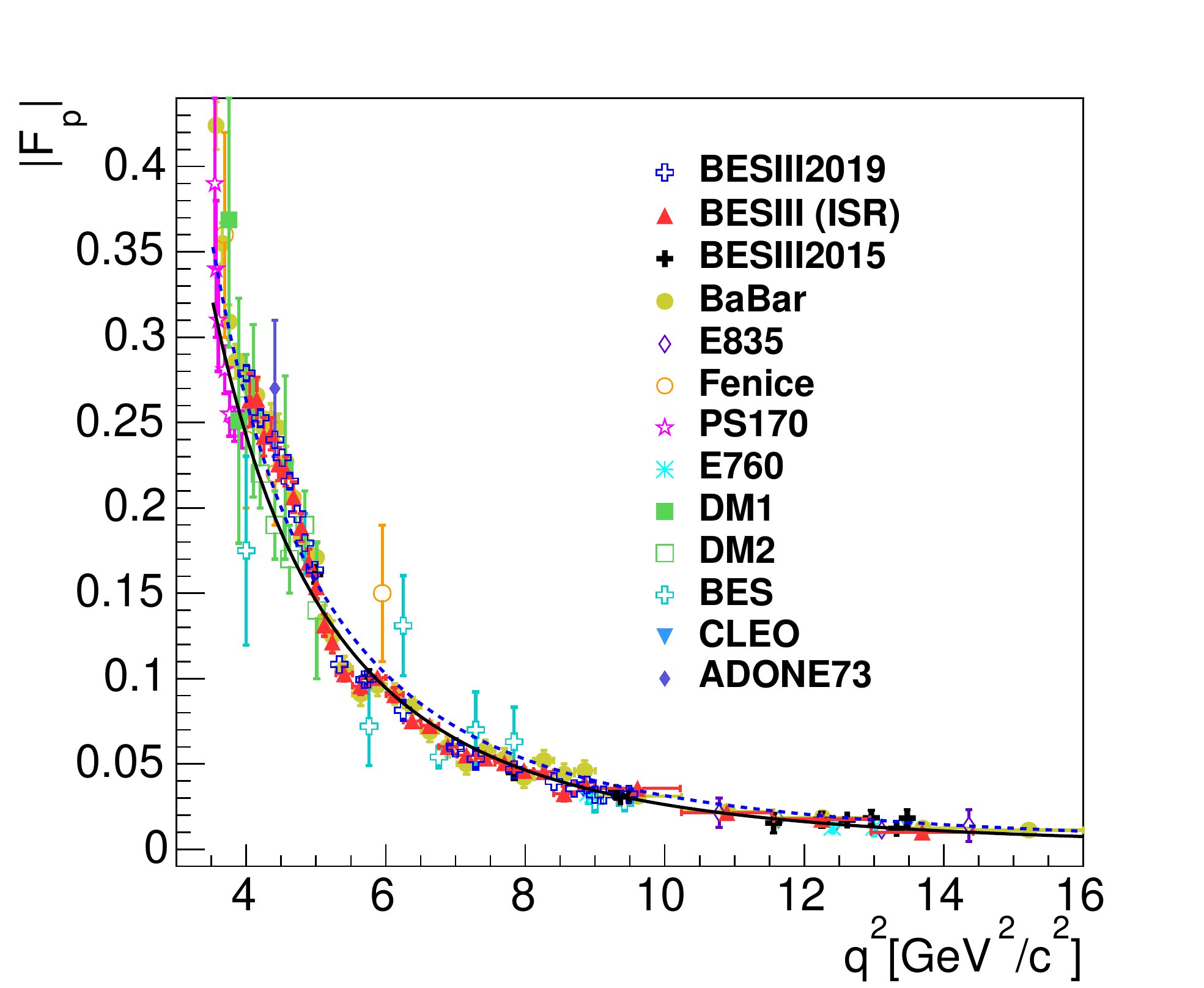}
\caption{The proton effective FF values measured  by: E835~\protect\cite{Andreotti:2003,Ambrogiani:1999}, Fenice~\protect\cite{Antonelli:1998fv}, PS170~\protect\cite{Bardin:1994am}, E760~\protect\cite{Armstrong:1993}, DM1~\protect\cite{Delcourt:1979}, DM2~\protect\cite{Bisello:1983,Bisello:1990}, BES~\protect\cite{Ablikim:2005}, BESIII~\protect\cite{Ablikim:2015,Ablikim:2019njl,Ablikim:2019eau}, CLEO~\protect\cite{Pedlar:2005}, {\it BABAR}~\protect\cite{Lees:2013b,Lees:2013},  and ADONE73~\protect\cite{Castellano:1973}. The blue dashed curve shows the QCD inspired parametrization~\cite{Shirkov:1997,Bian:2015} based on Eq.~(\ref{eqEffqcd}). The solid black curve shows the parametrization [Eq.~(\ref{rekalo})] suggested in Ref.~\cite{TomasiGustafsson:2001za}.}
\label{GEFFTOTPID}
\end{figure}

The world data on the proton effective FF are shown in Fig.~\ref{GEFFTOTPID}. Different parametrizations of the proton FFs can be found in literature \cite{Ambrogiani:1999,Brodsky:2007hb,TomasiGustafsson:2001za,Shirkov:1997}.  For example, the blue dashed curve in Fig.~\ref{GEFFTOTPID} represents the quantum chromodynamics (QCD) inspired $|F_{p}|$ parametrization from Refs.~\cite{Shirkov:1997,Bian:2015}:
\be
|F_{p}|=\frac{{\cal{A_{\rm QCD}}}}{q^4 [\log^2(q^2/\Lambda_{\rm QCD}^2)+\pi^2]},
\label{eqEffqcd}
\ee
where the parameters ${\cal{A_{\rm QCD}}}=72~(\mbox{GeV/$c$})^4$ and $\Lambda_{\rm QCD}=0.52~(\mbox{GeV/$c$})$ are obtained from a fit to the experimental data \cite{Bian:2015}. The data on the time-like effective FF can also be reproduced by the function proposed in Ref.~\cite{TomasiGustafsson:2001za},
\be
|F_{p}|=\frac{{\cal{A}}}{(1+q^2/m_a^2)[1+q^2/q_0^2 ]^2},
\label{rekalo}
\ee
where the fit parameters are ${\cal{A}}=22.5$, $m_a^2=3.6~(\mbox{GeV/$c$})^2$, and $q_0^2=0.71~(\mbox{GeV/$c$})^2$. This model is illustrated in Fig.~\ref{GEFFTOTPID} by the solid black curve. The two functions (Eqs.~\ref{eqEffqcd} and \ref{rekalo}) reproduce  the behavior of $|F_p|$  over a wide $q^2$ range. For the current studies, Eq.~\ref{rekalo} is used to parametrize the proton electric and magnetic FFs assuming their ratio, $R=|G_E|/|G_M|$, is equal to one. The region
between $q^2$ = 5.1  (GeV/$c$)$^2$ (laboratory beam momentum $p_{beam}$ = 1.5 GeV/$c$) and  8.2 (GeV/$c$)$^2$ ($p_{beam}$ = 3.3 GeV/$c$) where \PANDA  is expected to provide the first data on the time-like proton FFs using the process $\bar{p} p \rightarrow \mu^{+} \mu^{-}$ will be examined in this paper.

\begin{table*}[th!]
	\begin{center}
	\begin{tabular}{ l l l l l l l l }
	\hline
$p_{beam}$     &        $q^2$    &       ${\sigma}_{int}({\mu}^{+}{\mu}^{-})$  & $N_{int}({\mu}^{+} {\mu}^{-})$       &   ${\sigma}_{int}({\pi}^{+} {\pi}^{-})$   &     $N_{int} ({\pi}^{+} {\pi}^{-})$ &  $\displaystyle\frac{{\sigma}_{int}({\pi}^{+} {\pi}^{-})}{{\sigma}_{int}({\mu}^{+}{\mu}^{-})}$\\
 $[$GeV/$c$]      &         $[(\mbox{GeV/$c$})^{2}]$   &    $[$pb$]$    	           &                                                        &                             $[\mu$b$]$           &                                              &   x 10$^{-6}$   \\
	\hline
1.5                    &            5.08                        &                 641                                        &             128 x 10$^4$                              &                   133                                    &      265 x 10$^9$       &       0.21      \\                             
1.7                     &           5.40                        &                 415                                          &             830 x 10$^3$                         &                    101                                    &      202   x 10$^9$         &       0.24         \\
2.5                    &            6.77                       &                  89.2                                           &             178 x 10$^3$                         &                   22.6                                 &           452 x 10$^8$         &       0.25      \\
3.3                     &           8.20                        &                 24.8                                          &             497  x 10$^2$                          &                        2.96                                   &        593 x 10$^7$      &          0.12   \\
	\hline
	\end{tabular}
	\caption{Number of expected events $N_{int}$ and integrated cross-sections $\sigma_{int}$ based on Eq.~\ref{eq:B6} in the  |cos($\theta_{CM}$)| < 0.8 angular range. For the calculations, the FF parametrization of Eq.~\ref{rekalo} is used for the  $\bar{p} p \rightarrow {\mu}^{+} {\mu}^{-}$  signal reaction.  At 1.5 GeV/$c$ and 1.7 GeV/$c$, a cross-section is used for $\bar{p} p \rightarrow {\pi}^{+} {\pi}^{-}$  based on a fit of available data from \cite{Eisenhandler:1975kx} with Legendre polynomials. The 2.5 and 3.3 GeV/$c$  beam momentum values   correspond to the interpolation region of the pion cross section model. A time-integrated luminosity of $\mathcal{L}$ = 2 fb$^{-1}$ is assumed for each  $N_{int}({\mu}^{+} {\mu}^{-})$ and $N_{int}({\pi}^{+} {\pi}^{-})$  kinematical point.}

\label{tab:crosssections}
	\end{center}
\end{table*}

\section{Monte Carlo Simulation with PandaRoot}

\label{sec:pandaroot}

The offline software for the \PANDA detector simulation and analysis,  PandaRoot \cite{Spataro:2012},  has been  developed within the framework for the future FAIR experiments, FairRoot \cite{fairroot}. The PandaRoot software encompasses full detector simulation and event reconstruction. In order to investigate the feasibility to use the $\bar{p} p \rightarrow \mu^{+} \mu^{-}$ process for the measurement of the proton time-like FFs at \PANDA, Monte Carlo (MC) simulation studies based on PandaRoot are performed.

\subsection{Generation of the  $\bar{p} p \rightarrow {\mu}^{+} {\mu}^{-}$ signal events}

\begin{figure}
    \centering
\resizebox{0.5\textwidth}{!}{
  \includegraphics{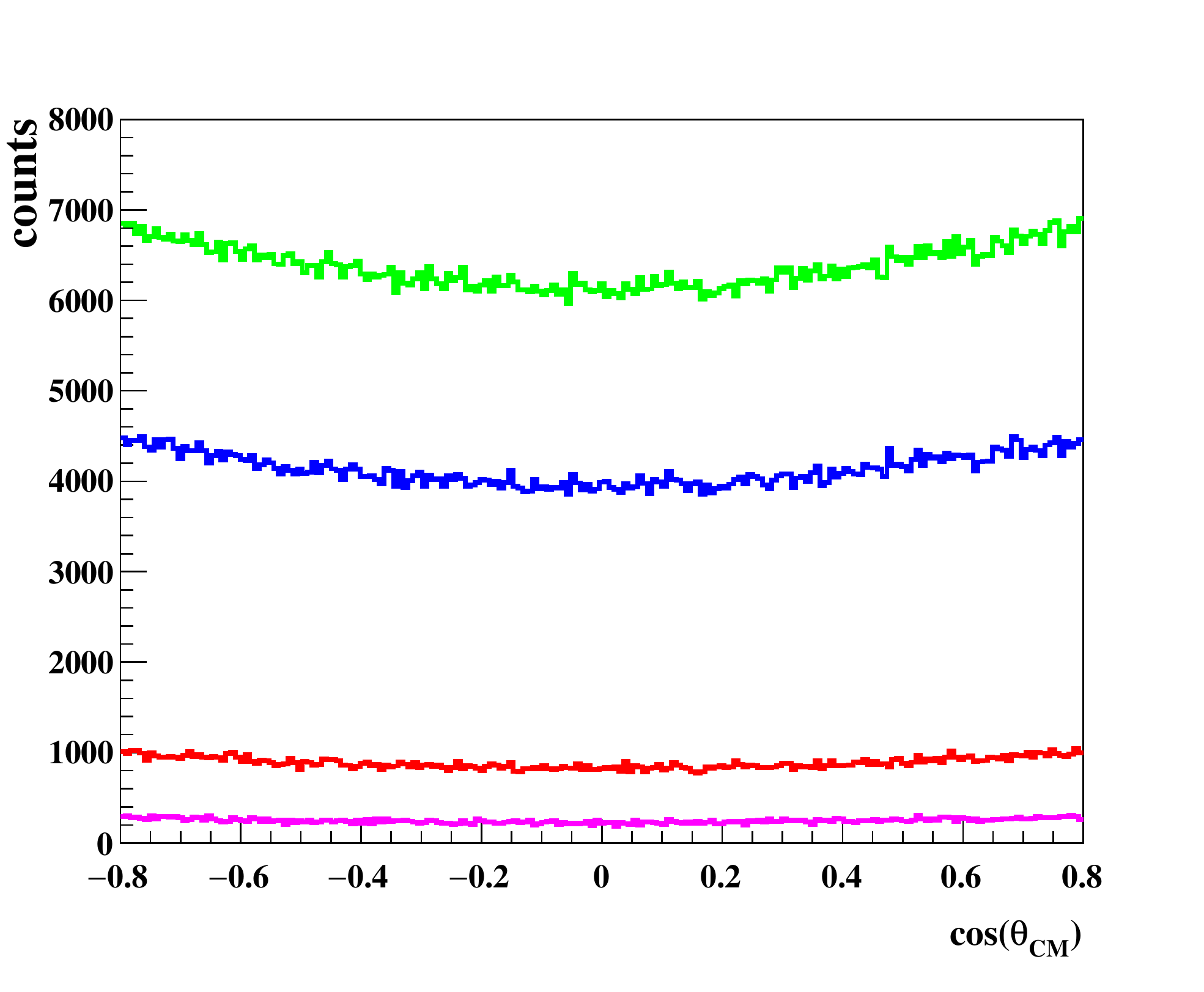}
}
\caption{Angular distribution of the MC generated $\mu^-$ in the $\bar{p}p$ CM frame for the $\bar{p}p \rightarrow {\mu}^+{\mu}^-$ process (sample S2) for $p_{beam}$ = 1.5 GeV/$c$ (green), 1.7 GeV/$c$ (blue), 2.5 GeV/$c$ (red) and 3.3 GeV/$c$ (magenta) (color online). Due to the \PANDA detector acceptance the events are generated in the  |cos(${\theta}_{CM}$)|< 0.8 angular range.}
\label{fig:signal}     
\end{figure}

The signal event generation at different beam momenta values, $p_{beam} \in$ (1.5, 1.7, 2.5, 3.3) GeV/$c$, is based on the expression of the differential cross-section (Eq. \ref{eq:B6}) as a function of the time-like electromagnetic proton FFs. Equation \ref{rekalo} is used for the parametrization of $|G_{E}|$ and $|G_{M}|$. For each value of beam momentum, the number of expected signal events is extracted (see Tab. \ref{tab:crosssections}), assuming a time-integrated luminosity of 2~fb$^{-1}$. The MC sample denoted as S1 is produced at each value of the beam momentum with a large amount of events to determine the signal efficiency; as a result, the statistical uncertainty on the efficiency is negligible. Additional MC samples (S2) are generated based on the numbers of expected signal events for a proper consideration of the statistical fluctuations and uncertainties. The samples S2 represent the signal events that will be collected in the future at the \PANDA experiment. The dependence of the expected number of signal events on cos($\theta_{CM}$) is illustrated  in Fig. \ref{fig:signal} at beam momenta of 1.5 GeV/$c$, 1.7 GeV/$c$, 2.5 GeV/$c$, and 3.3 GeV/$c$.

\subsection{Largest background sources}

 The suppression of the hadronic background is one of the main experimental challenges for the measurement of the time-like proton FFs in the muon channel.  The possible background channels with the largest cross sections (see also \cite{Boucher:2011,Sudol:2010,AlaasThesis,Wang:2015ybw,Wang:2017pkc}) are
  
  \begin{itemize}
    \item $\bar{p}p \rightarrow \pi^+\pi^-$;
      \item $\bar{p}p \rightarrow K^+K^-$;
                \item $\bar{p}p \rightarrow K^+K^-{\pi}^0$;
                 \item $\bar{p}p \rightarrow \pi^+\pi^-{\pi}^0$;
                 \item $\bar{p}p \rightarrow \pi^+\pi^-\omega$;
        \item $\bar{p}p \rightarrow \pi^+\pi^-{\rho}^0$;
       \item $\bar{p}p \rightarrow n\pi^+ n\pi^- m{\pi}^0$ with n $\geq$ 2 and m $\geq$ 0.
  \end{itemize}
  
The main background source is the production of two charged pions ($\bar{p} p \rightarrow \pi^{+} \pi^{-}$). Its total cross section is estimated to be a factor of $10^{5}$-$10^{6}$ larger than the signal, depending on the beam energy \cite{Eisenhandler:1975kx,VandeWiele:2010kz,Wang:2015ybw}. Therefore, an efficient background suppression together with a sufficient signal efficiency is required to extract the desired signal. Furthermore, due to their similar masses, it is difficult to distinguish between muons and pions, especially at higher particle momenta where they show a quite similar behavior inside the \PANDA MS.

\subsection{$\bar{p}p \rightarrow \pi^{+}\pi^{-}$ background generation }

\begin{figure}
    \centering
\resizebox{0.5\textwidth}{!}{
  \includegraphics{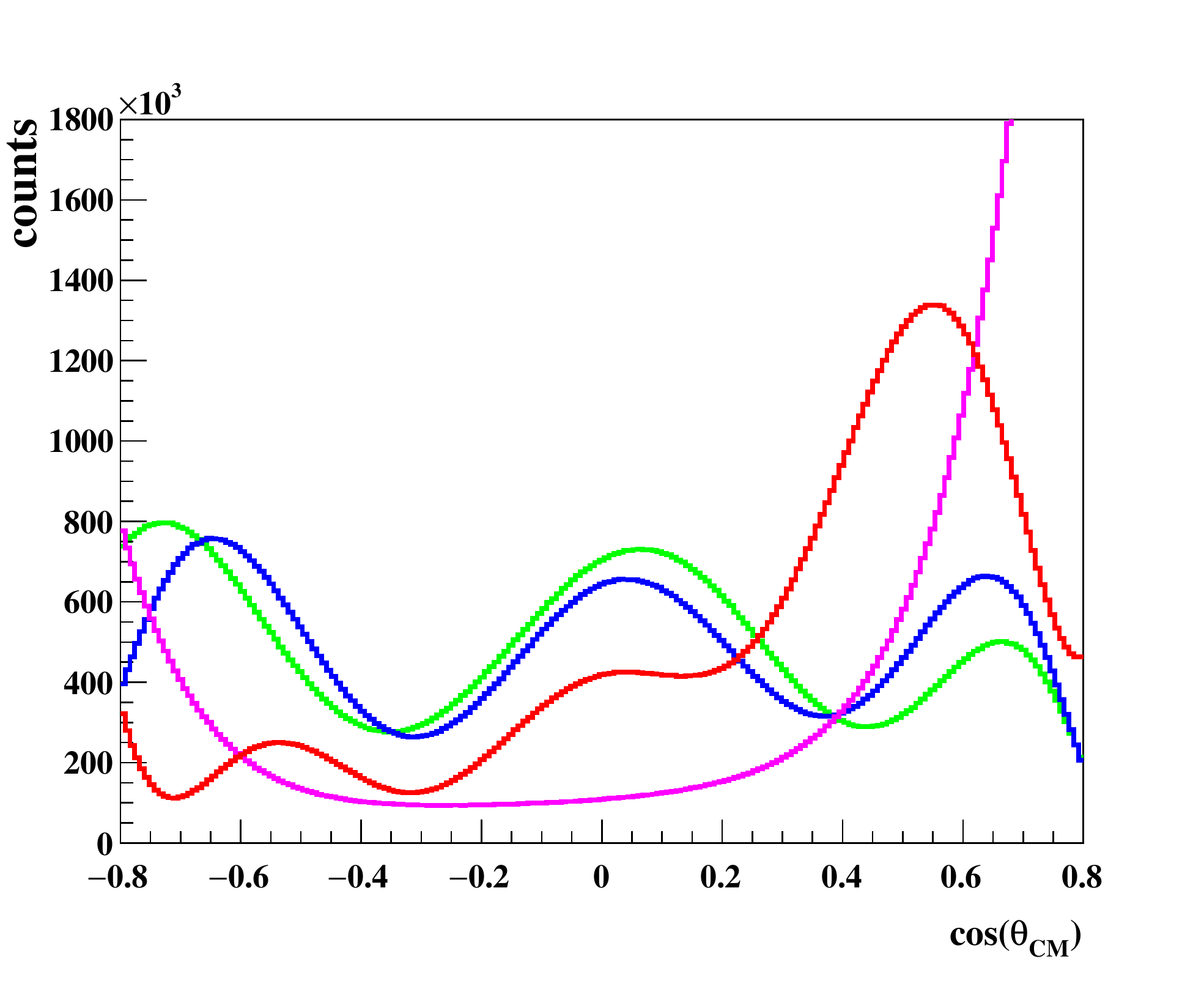}
}
\caption{Angular distribution of the generated $\pi^{-}$ in the $\bar{p}p$ CM frame for the $\bar{p}p \rightarrow {\pi}^+{\pi}^-$ process (sample B1) for $p_{beam}$ = 1.5 GeV/$c$ ({\textbf{green}}), 1.7 GeV/$c$ ({\textbf{blue}}), 2.5 GeV/$c$ ({\textbf{red}}) and 3.3 GeV/$c$ ({\textbf{magenta}}).}
\label{fig:backgrounddistribution}     
\end{figure}

For the simulation of the main background channel $\bar{p}p \rightarrow \pi^{+}\pi^{-}$ (referred in the following as pion background), a dedicated event generator has been developed  \cite{Zambrana:2014} based on two phenomenological parameterizations in different beam momentum ranges. For antiproton  momenta  in the   $0.79\leq p_{beam}\leq 2.43$ GeV/$c$ range, the \grqq low energy region\grqq, a combination of Legendre polynomials  reproduces the data well and is  used to fit the available data (see \cite{Eisenhandler:1975kx}). The oscillating behavior of the angular distribution at lower momentum (see Fig. \ref{fig:backgrounddistribution}) is due to contributions of higher L waves in the relative motion in the di-pion system.

For  $5.0 \leq p_{beam} \leq 12.0$ GeV/$c$ beam momenta, the so-called \grqq high energy region\grqq, a Regge-inspired parametrization from \cite{VandeWiele:2010kz} is tuned on the data from \cite{Eide:1973tb,Buran:1976wc,Armstrong:1986ng,White:1994tj}.  The angular distribution of the $\pi^-$ loses its oscillating behavior, becoming  forward or backward peaked. This corresponds to small values of the Mandelstam variables \textit{t} or \textit{u}, respectively, to which different exchange particles contribute.

For momenta in the  $2.43 < p_{beam} < 5.0$ GeV/$c$ intermediate region, an interpolation is used since there are no available data or valid models providing a reliable description.

A data sample (B1)  consisting  of $10^{8}$ background events, is simulated in the  $|cos(\theta_{CM})| < 0.8$ range at each value of the beam momentum. The angular distribution of the generated $\pi^{-}$ is depicted in Fig. \ref{fig:backgrounddistribution} at $p_{beam}$ = 1.5 GeV/$c$ (green), 1.7 GeV/$c$ (blue), 2.5 GeV/$c$ (red) and 3.3 GeV/$c$ (magenta).

\section{Analysis of the simulated event samples}

After event generation, the particles are propagated through the material of the \PANDA detector using the GEANT4 software package.  The digitization of the analog detector signals is then simulated, followed by the event reconstruction. At this step the reconstruction of the trajectories in the sub-detectors is done by fitting the charged particle tracks. Finally, particle identification and the analysis of the reconstructed data are performed.

\subsection{Event reconstruction}

The events for the signal and background reactions are reconstructed based on the following conditions:
\begin{itemize}
\item events with at least one  positive and one negative track are selected. If more than one positive-negative track pair can be combined, the pair with the ${(\theta^{+}+\theta^{-})}_{CM}$ closest to 180$^\circ$ is selected; 
\item both particle candidates must have at least one hit each in the MS.
\end{itemize}

\subsection{Kinematic and PID variable information}
\begin{figure*}
\centering
\begin{subfigure}{.5\textwidth}
  \centering
  \includegraphics[width=0.9\textwidth]{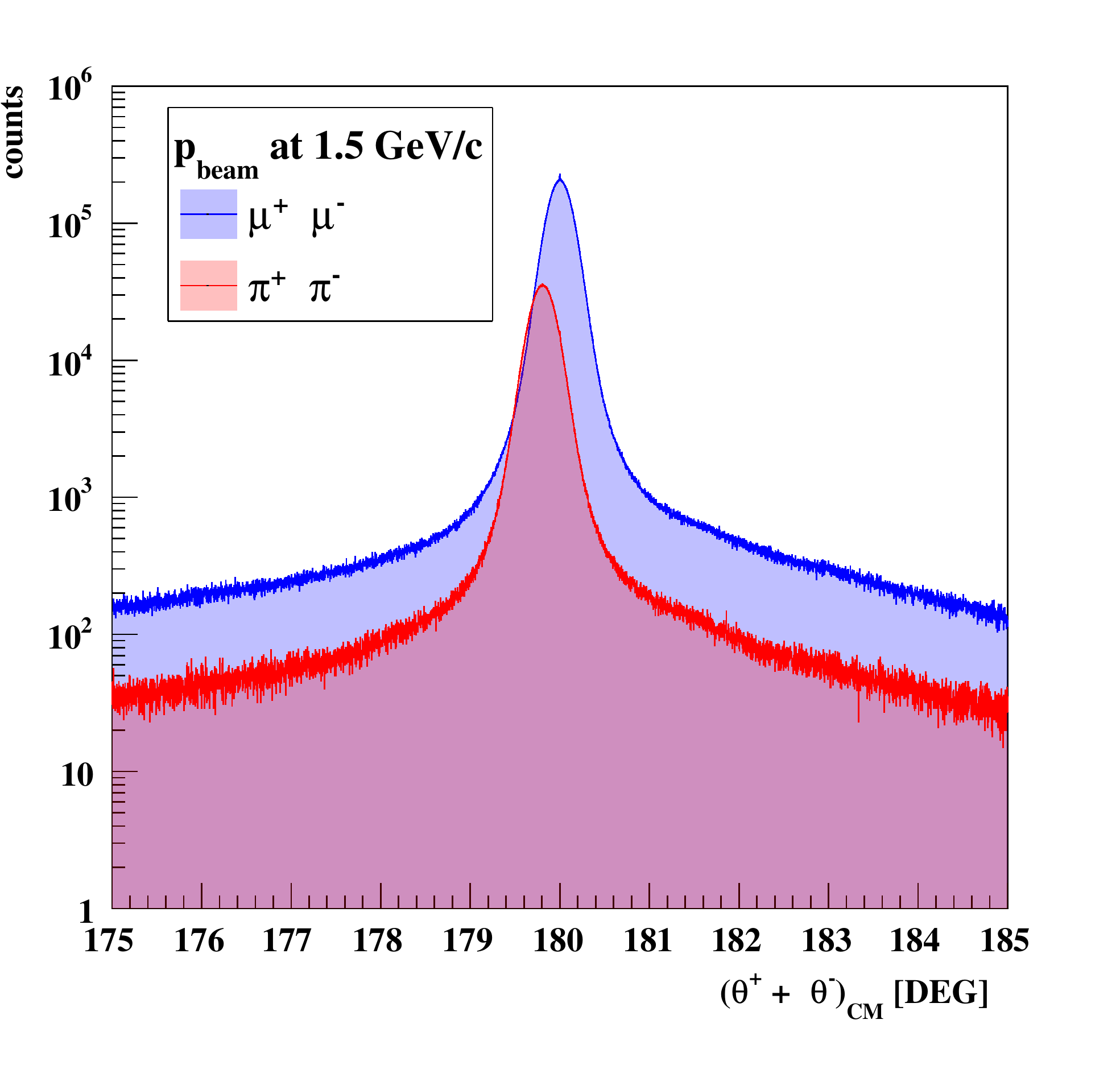}
     \caption{}
  \label{fig:Thetasum_1_5}
\end{subfigure}%
\begin{subfigure}{.5\textwidth}
  \centering
  \includegraphics[width=0.9\linewidth]{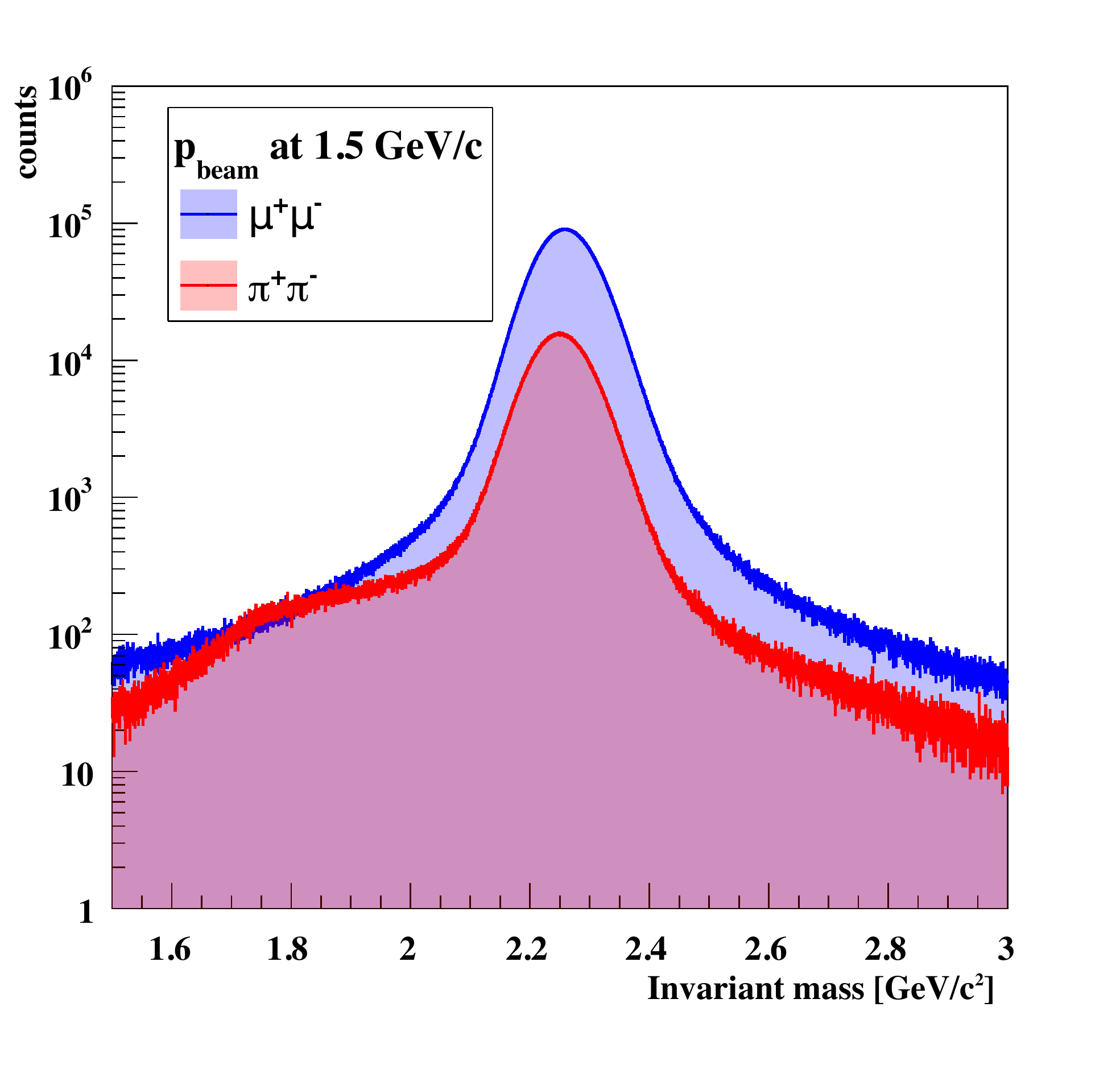}
     \caption{}
  \label{fig:Invmass_1_5}
\end{subfigure} \\
\begin{subfigure}{.5\textwidth}
  \centering
  \includegraphics[width=0.9\textwidth]{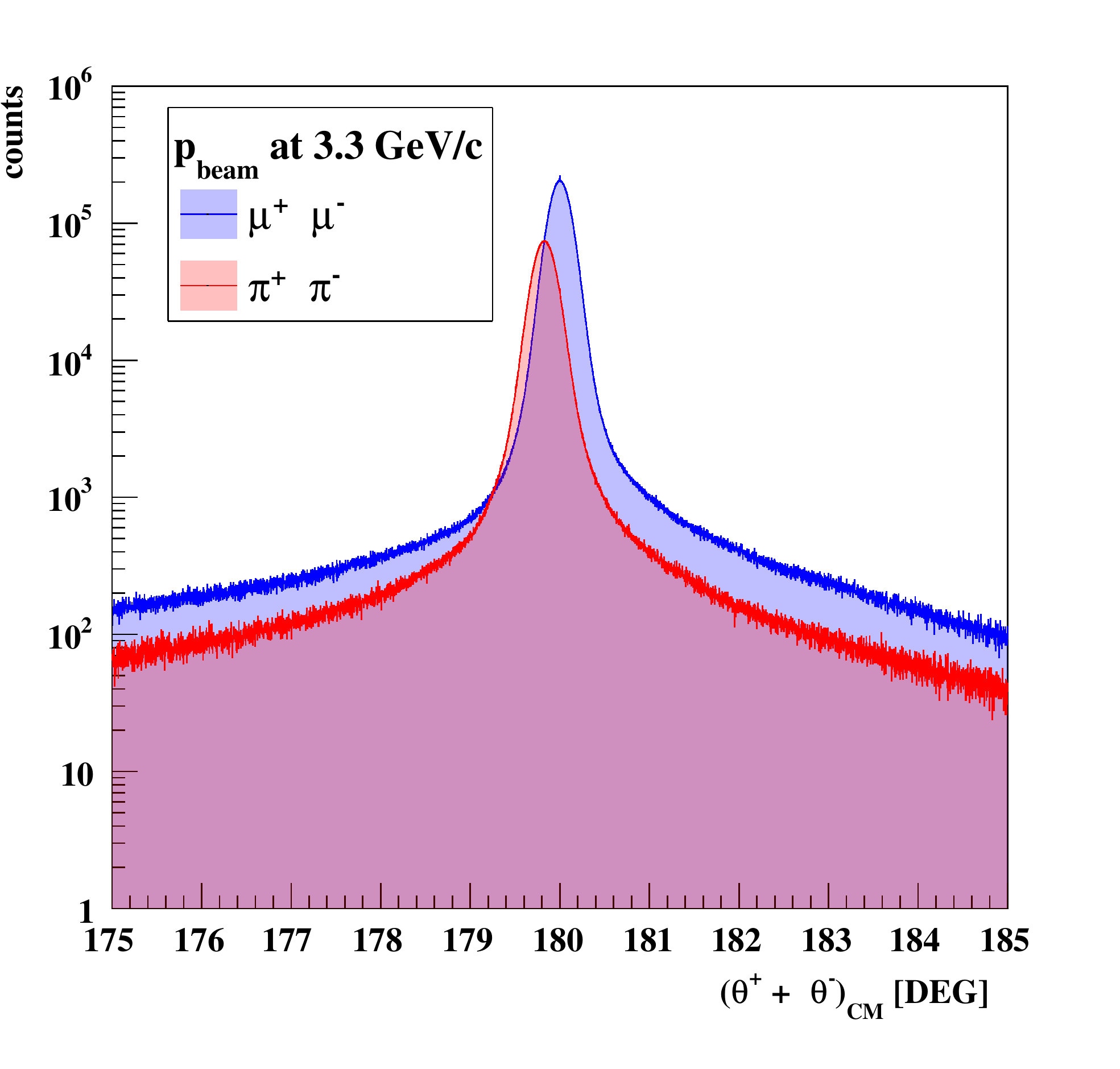}
     \caption{}
  \label{fig:Thetasum_3_3}
\end{subfigure}%
\begin{subfigure}{.5\textwidth}
  \centering
  \includegraphics[width=0.9\linewidth]{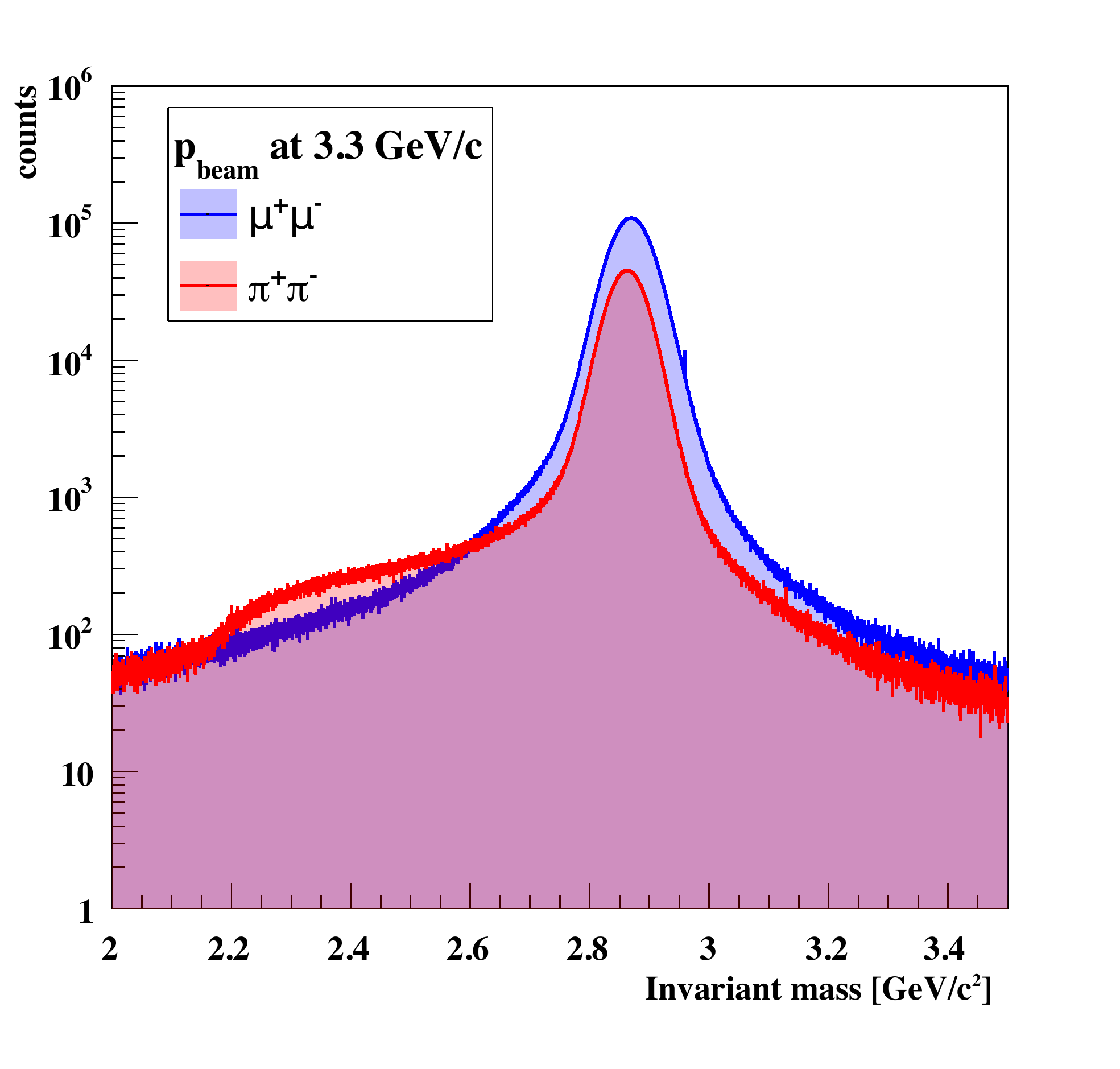}
     \caption{}
  \label{fig:Invmass_3_3}
  \end{subfigure}
\caption{Left column: Distribution of the  kinematic variable ${(\theta^{+}+\theta^{-})}_{CM}$   for the signal (blue) and the background (red) reconstructed events. The plots describe the S1 and B1 samples at (a) $p_{beam}$ = 1.5 GeV/$c$ and (c)  $p_{beam} = $3.3 GeV/$c$. Due to the muon mass hypothesis, the peak of the background distribution is shifted to slightly  smaller angles. Right column: Distribution of the invariant mass $M_{inv}$ of the particles final state  for the signal (blue) and the background reconstructed events (red) at (b) $p_{beam} = $1.5 GeV/$c$ and (d)  $p_{beam} = $3.3 GeV/$c$.}
\label{fig:2}       
\end{figure*}

Kinematic selections can be used to suppress contributions from hadronic channels with more than two particles in the final states, as well as events with secondary particles originating in the interaction with the detector materials. One of the kinematical variables is the sum of the polar angles of both charged tracks in the CM frame.  The angles are derived from  the three-momenta at the vertex, which are based on the reconstructed trajectory using the information of both STT and MVD.  The particle's energy at the production vertex is calculated assuming the muon mass hypothesis. The total polar production angle is depicted in Fig. \ref{fig:2} (a,~c). One can see that the peak of the background distribution is shifted to smaller angles in comparison to the signal peak, due to assigning the background pions the muon mass hypothesis. Therefore this variable can be used for the signal-background separation. In addition, both tracks are ideally emitted back-to-back in the lab frame in a plane perpendicular to the beam so the azimuthal angle difference  ${(|\phi^{+}-\phi^{-}|)}_{lab}$ is ideally peaked at 180$^\circ$. From the 4-momenta of both tracks, the invariant mass is calculated:
\begin{equation}
\label{eq:Minv}
M_{inv}=\sqrt{{(p_{\ell^+}+p_{\ell^-})}^2}.
\end{equation}
The corresponding $M_{inv}$ distributions are shown in Fig. \ref{fig:2} (b,~d). The invariant mass spectrum shows a hump at the region around 1.8 GeV/$c^2$ at $p_{beam}$ = 1.5
GeV/$c$ and 1.7 GeV/$c$, which is caused by the decay of a single pion ($\pi \rightarrow \mu \nu$). For $p_{beam}$ = 2.5 GeV/$c$ and 3.3 GeV/$c$, this hump starts around 2.0 GeV/$c^2$ and 2.2 GeV/$c^2$, respectively, due to the higher beam momentum.

The most important subdetector for the $\mu$/$\pi$ separation is the MS. Its sandwich structure consists of alternating active and passive layers, which allow the different behavior of $\mu$ and $\pi$ inside the detector to be distinguished. Pions interact via both ionization energy loss and hadronic showering, while muons interact only through ionization.

A highly energetic pion is misidentified as a muon when a) it undergoes only ionization processes inside the MS material and b) it decays into a muon and the corresponding (anti-)neutrino. After the $\mu$-selection, only muons from pion decay can enter in the pionic background.

The momenta of the produced particles decrease with increasing values of ${\theta}_{CM}$. Hence, most of the particles are absorbed by the MS at large backward angles ${\theta}_{CM}$. As an example, at $p_{beam}$ = 1.5 GeV/$c$ this behavior can be seen at angles approximately bigger than 100 degrees. Particles which are produced under smaller angles are able to cross through the MS due to their higher momenta.

\begin{figure*}[t]
\begin{center}
\resizebox{0.90\textwidth}{!}{
 \includegraphics{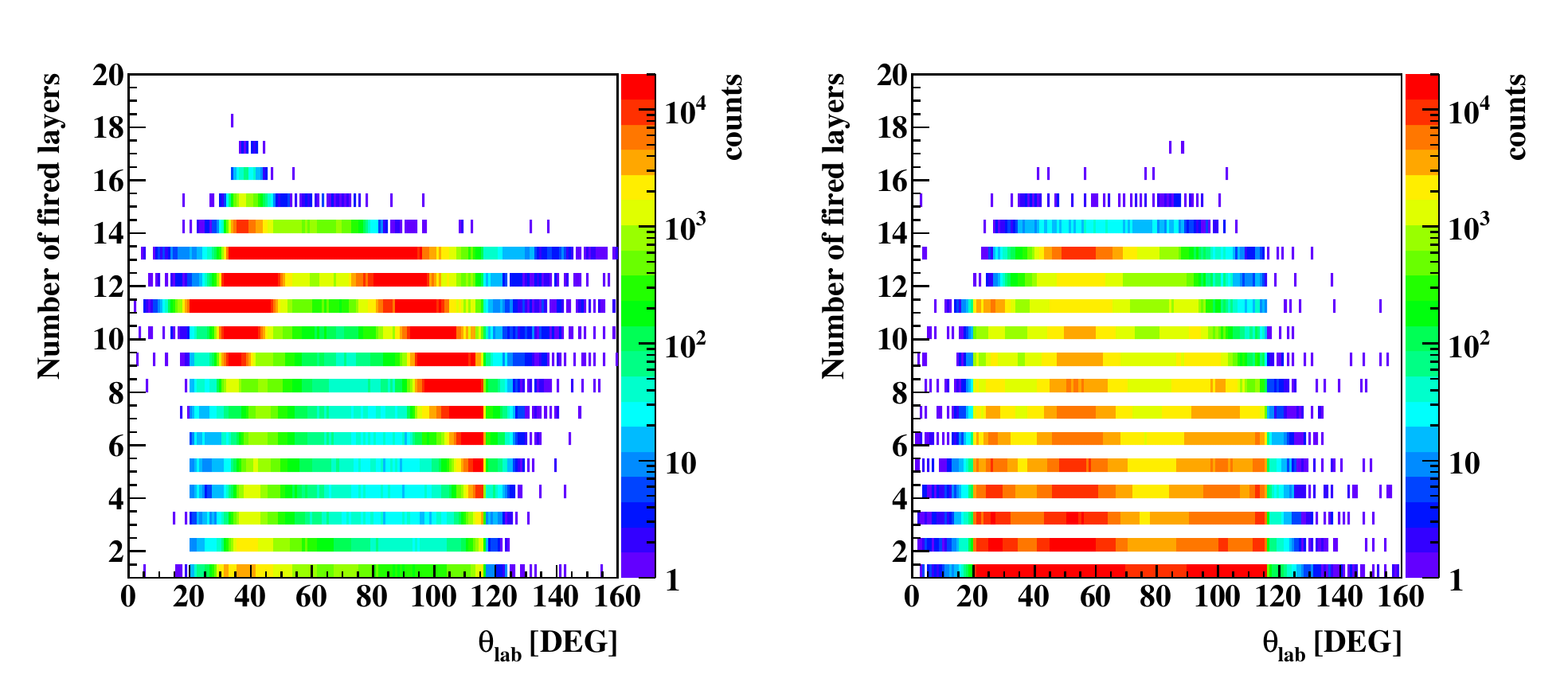}}
\caption{Angular dependence of the number of fired detection layers in the MS for negatively charged particles for the signal (left column) and the background (right column) reconstructed events at $p_{beam}$ = 1.5 GeV/$c$. The different behavior of muons and pions in the MS is crucial for an efficient $\mu/\pi$ separation. From the differences in the detector response, one can deduce that the number of fired detection layers has a strong separation power for muons and pions.}
\label{fig:NoFLay}
\end{center}
\end{figure*}

\begin{figure*}[t]
\begin{center}
\resizebox{0.90\textwidth}{!}{
 \includegraphics{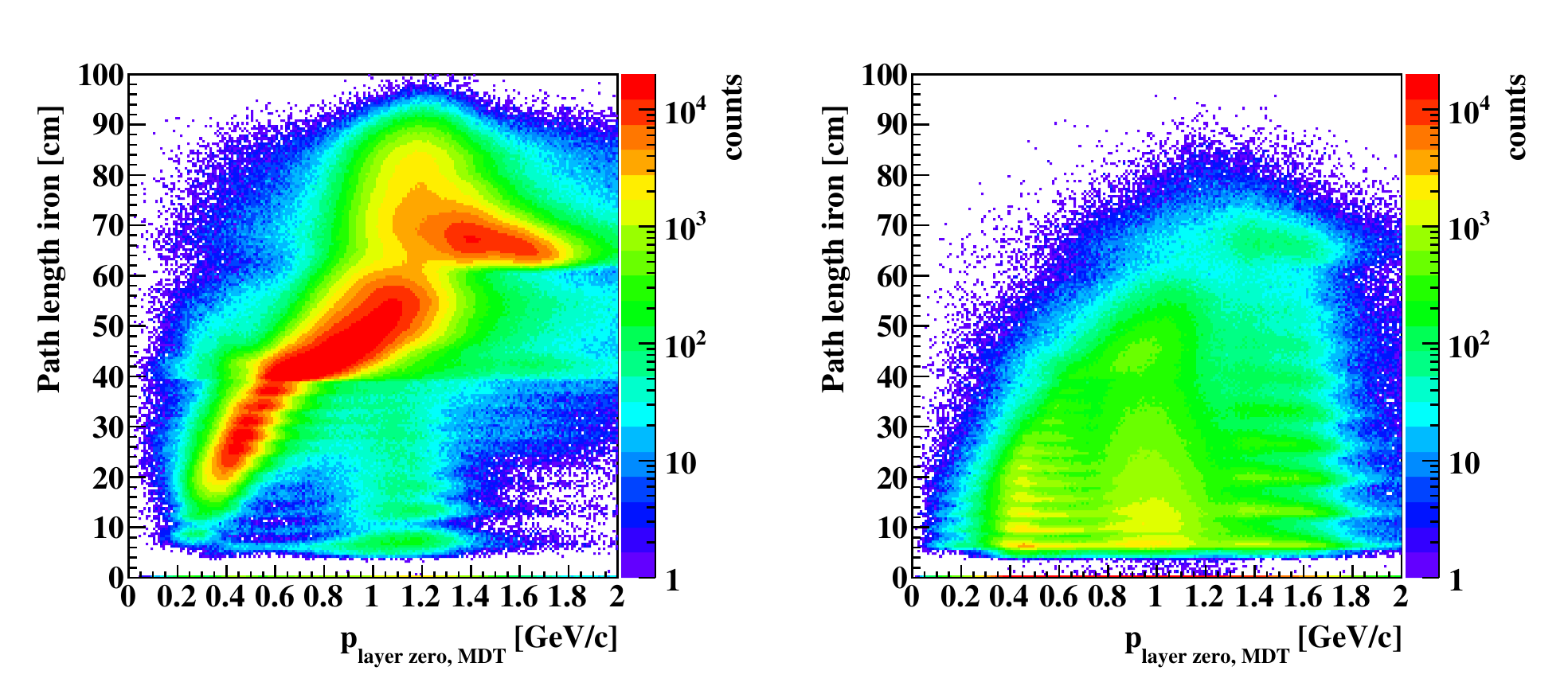}}
\caption{Dependence of the path length in the iron absorber of the MS on the incident particle momentum , for negatively charged particles for the signal (left column) and background (right column) reconstructed events at $p_{beam}$ = 1.5 GeV/$c$. This variable has a strong separation power and is used, in combination with $p_{layer zero, MDT}$, for the determination of the identification probability for muons, named P($\mu$), by the MS.}
\label{fig:irondepth}
\end{center}
\end{figure*}

Figure \ref{fig:NoFLay} shows the number of fired detection layers versus the reconstructed polar angle for negatively charged tracks from the signal (left column) and the background (right column). Most of the $\pi^-$'s from the pion background sample are absorbed within the first layers of the MS at $p_{beam}$ = 1.5 GeV/$c$. At this momentum, about 7$\%$ of all the tracks cross 11 detection layers or more. Of that fraction, the MC truth information shows that about 9$\%$ are decay muons ($\mu^{-}$) and the rest of the particles are $\pi^{-}$. A negligible fraction (at the level of a few per mille) consists of misidentified particles of opposite charge ($\mu^{+}$ and  $\pi^{+}$).

Another important variable for signal-background separation is the path length of the trajectories inside the MS. The dependence of this path length on the track momentum at the entrance of the MS is shown in Fig. \ref{fig:irondepth} for reconstructed negatively charged particles in the high statistics signal samples (left) and the background sample (right). Note that the expected path length for a given incident particle momentum is strongly correlated with the incident polar angle.

The observables measured in the STT and EMC are generally less powerful for mu/pi separation, as both detectors respond in a very similar way to these particles. The relevant variables are the deposited energy inside the electromagnetic calorimeter ($E_{EMC}$/$p$, with $p$  the magnitude of the reconstructed particle 3-momentum at the interaction vertex), the EMC lateral moment and the mean energy loss per unit of length in the Straw Tube Tracker ($dE/dx$ ${STT}$). The EMC lateral moment is defined as:
\begin{equation}
\label{eq:latmom}
LAT = \frac{\sum_{i=3}^{N} {E}_{i}{r_{i}}^{2}}{\sum_{i=3}^{N} {E}_{i}{{r}_{i}}^2+{E}_{1}{r_{0}}^{2}+{E}_{2}{{r}_{0}}^{2}},
\end{equation}
where N is the number of crystals hit by the shower and $E_{i}$ is the deposited energy in the i-th crystal in the shower, with $E_1$ > $E_2$ > ... > $E_N$. The lateral distance between the central and the i-th crystal is given by $r_i$.
Here $r_0$ stands for the fixed average distance between two crystals. Since the numerator does not contain the three highest energy depositions, the ratio will be smaller  for electromagnetic showers in comparison to  hadronic showers.

Despite the low separation power of the EMC and STT variables by themselves, they can help to improve the signal-background separation when multivariate data classification is used to optimize the $\mu/\pi$ separation.

The identification probability for being a muon, named P($\mu$), is determined based on two variables from the MS; the path length inside the iron absorber of the MS, denoted as the iron depth, and the initial particle momentum p$_{layer zero, MDT}$ (= p$_{lMDT}$) measured at the detector entrance. Threshold values are defined for both of them and depend on the MS module. Further studies based on measurements using a real muon system prototype are planned and will help to achieve a marginal improvement of the PID capability for muons.

\subsection{Optimizing the $\mu$/$\pi$ separation by using Boosted Decision Trees}

The analysis of the simulated data aims to achieve the best possible background suppression while keeping a sufficient signal efficiency. Multivariate data classification is used to optimize the signal-background separation. Signal efficiency and background suppression studies are based on the high statistics $\mu^{+}\mu^{-}$ sample (S1) and the  high statistics $\pi^{+}\pi^{-}$  sample (B1). After the event reconstruction, the full analysis based on multivariate data analysis (MVA) is carried out. Different methods of multivariate data classification are investigated using the Root-integrated software package Toolkit For Multivariate Data Analysis with ROOT (TMVA) \cite{root,tmva}.

For MVA, a set of input variables is needed. The most important detector related to the $\mu$/$\pi$ separation is the Muon Range System (MS), as discussed in the previous sector.

\subsubsection{MVA analysis}
To summarize, for the analysis the following input variables are considered:

\begin{itemize}
\item the path length inside the iron absorber of the MS, denoted as "iron depth";
\item the number of fired layers in the MS; 
\item the initial momentum at MS layer zero: p$_{layer zero, MDT}$;
\item the normalized path length of the tracklet inside the MS to p$_{layer zero, MDT}$;
\item the identification probability for being a muon based on MS observables: P($\mu$);
\item the ratio of the deposited energy inside the EMC to the reconstructed momentum of the associated track: E$_{EMC}$/p;
\item the lateral moment of the EMC;
\item the deposited energy inside  a 3x3 crystal clusters, the central cluster being defined by the  maximum energy deposition;
\item the mean energy loss per unit of length inside STT, ${(dE/dx)}_{STT}$;
\item the number of hits inside STT.
\end{itemize}

Kinematical variables are used as input variables as well, although they are, generally less powerful, and are mainly used for data selection after MVA:
\begin{itemize}
\item the sum of the polar angles in the CM system: ${({\theta}^{+}+{\theta}^{-})}_{CM}$;

\item the invariant mass of the final state particles: $M_{inv}$. 

\end{itemize}
The following "spectator variables" are not used for the training, but are stored into the output tree together with the response of the multivariate classifiers:
\begin{itemize}
\item the azimuthal angle  ${(|\phi^{+}-\phi^{-}|)}_{lab}$ difference;
\item the CM polar angle of the negative final state particle: cos(${\theta_{CM}}$);
\item the CM polar angle of the positive final state particle.
\end{itemize}

Two different data sets feed the selected classifiers, both containing the reconstructed events together with the MC truth information. For the training, the classifiers use 50$\%$ of the input events, the remaining amount serving as test data for the trained classifiers. On the basis of these studies, different classification methods like Fisher Discriminants, Neural Networks or Boosted Decision Trees are trained, tested and their separation performance evaluated. The trained classifiers are stored as weight files and can be used afterwards to classify sets of unknown data.
 A helpful criterion to evaluate the performance of the different classifiers is the Receiver Operating Characteristics curve (ROC curve) which shows the achievable background rejection as a function of the corresponding signal efficiency.

\begin{figure}
    \centering
\resizebox{0.5\textwidth}{!}{
  \includegraphics{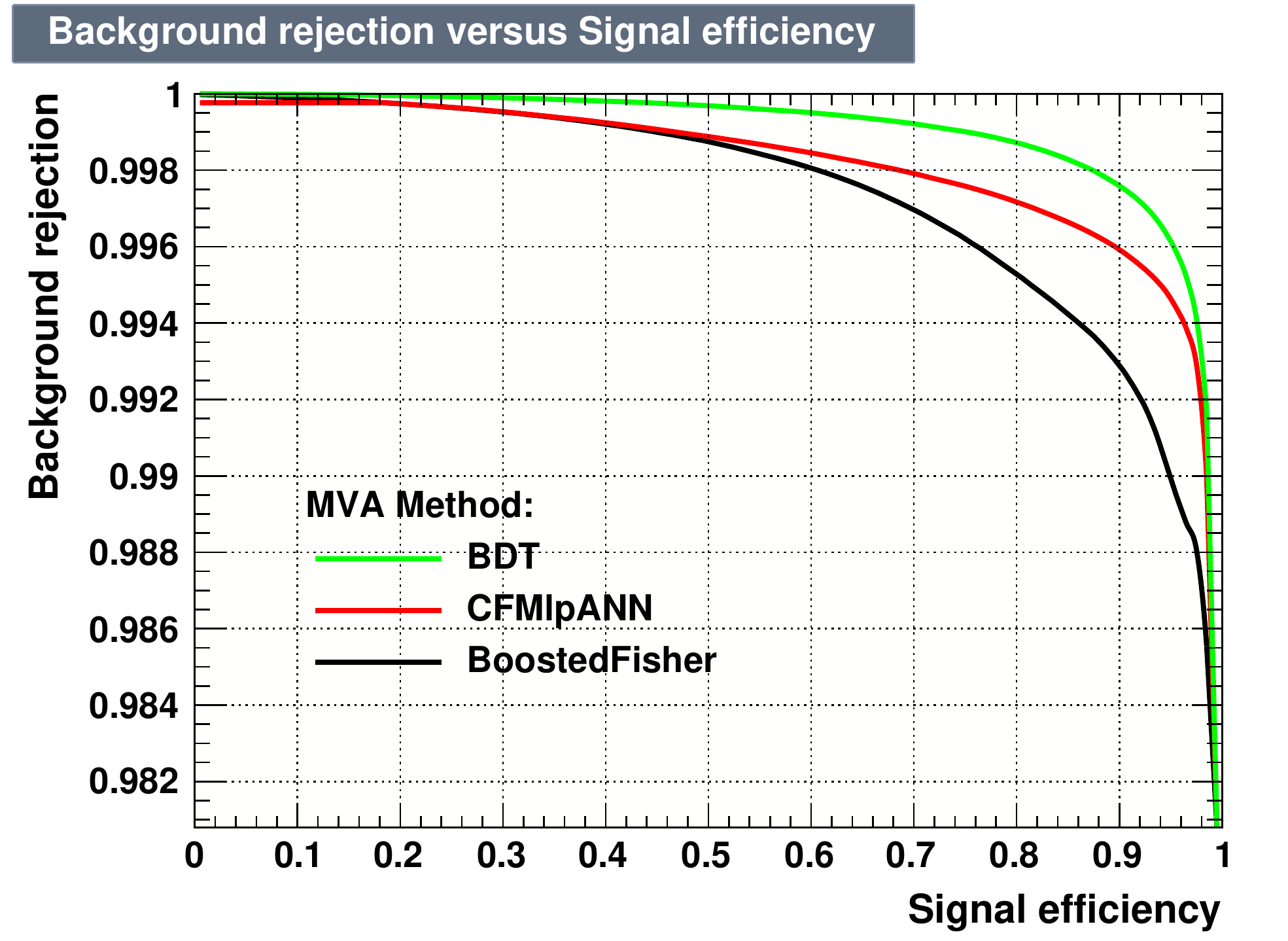}
}
\caption{The classifiers ROC curves deliver information on the performance of each classification method (for $p_{beam}$ = 1.5 GeV/$c$). The bigger the area underneath each curve, the higher is the quality of the classifiers performance. Here the performance of boosted decision trees (BDT) is compared to an artificial neural network (CFMlpANN) and a method based on boosted Fisher discriminants. The curves in this Figure are based on training data samples for signal and background containing 2*$10^5$ events.}
\label{fig:ROC}     
\end{figure}

In this work, Boosted Decision Trees show the best performance for $\mu$/$\pi$ separation. Fig. \ref{fig:ROC} reports the performance of different multivariate classifiers, which are applied on signal and background data sets after reconstruction, which have been generated only for the training of the classifiers. Each of those data samples contains 2*$10^{5}$ events. The area below each classifier curve can be used to judge on the quality of the classifiers performance. A high signal purity demands a very high background rejection which on the other hand implies small signal efficiencies.

\begin{figure}
    \centering
\resizebox{0.5\textwidth}{!}{
  \includegraphics{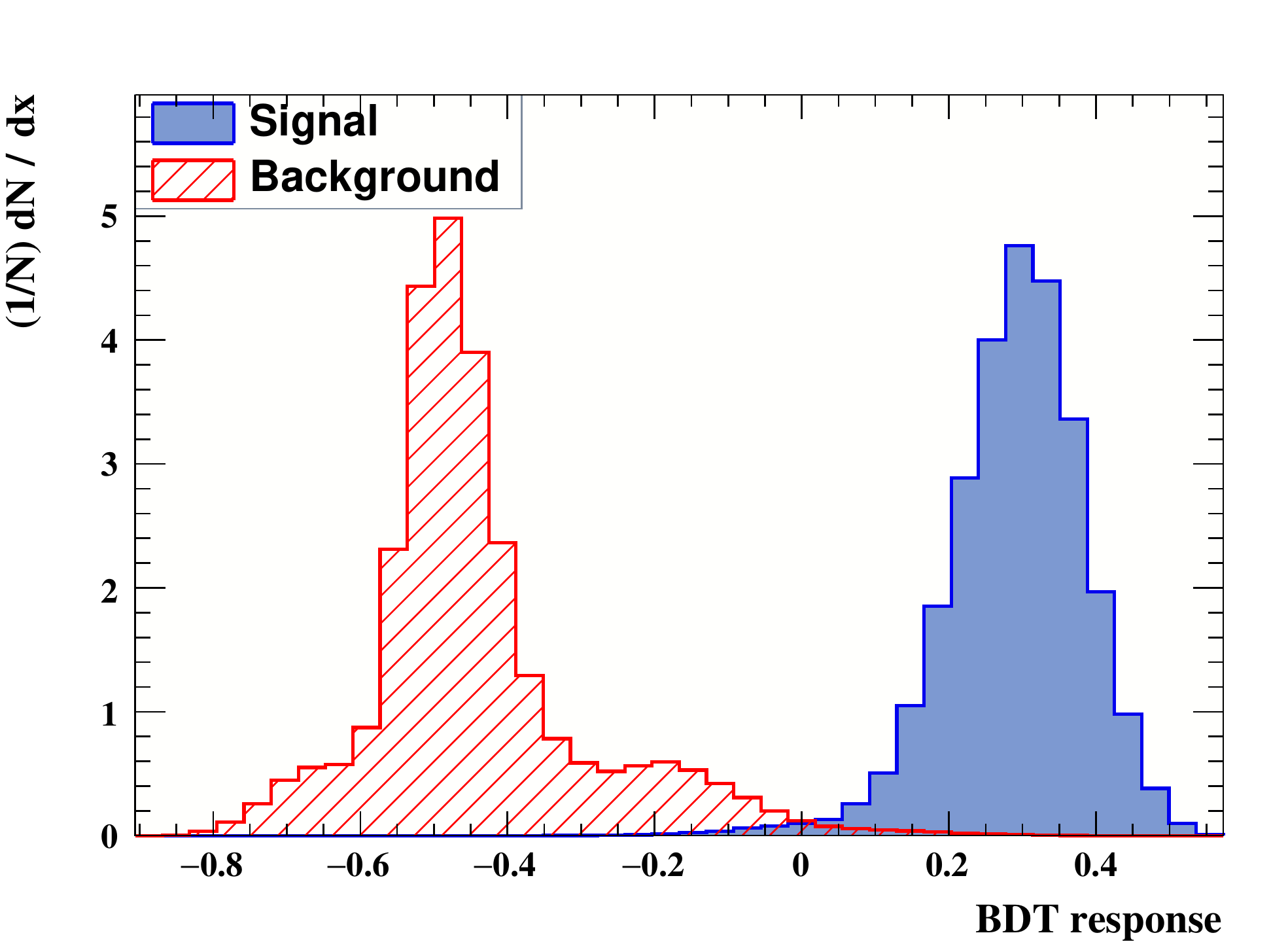}
}
\caption{Output distributions of the boosted decision tree classifier using adaptive boosting ("BDT response") after the training for the signal events (blue) and the background events (red) from the training data sample (2 * $10^5$ events each). Due to the overlap of both distributions it is not possible to fully reject the background while keeping an acceptable signal efficiency.}
\label{fig:BDT}     
\end{figure}

The BDT response for the signal and background samples is shown in Fig. \ref{fig:BDT}.  A classifier must be checked for overtraining, to reject cases of overfitting the classifier parameters to statistical fluctuations in the training data set. TMVA does this by comparing the event distributions from the training data and the values predicted by the classifiers. If the event selections ($2.1 < M_{inv} < 2.4 $ GeV/c$^2$) as well as the ${({\theta^{+}}+{\theta^{-}})}_{CM}>178.0^\circ$ cut are applied to the training data, the quality of the training data and the classification performance is improved. In the current studies an amount of $2*10^5$ events used in the TMVA procedure produced the best results. Half of the event sample is used to train the classifier, and the other half are used for testing.

The event selections considered in this work are summarized in Tab. \ref{tab:selection}. Selections are applied on the BDT outputs and on the distributions of the kinematical variables.

The signal and background efficiencies are mainly affected by the cuts on the BDT values. Sufficient signal statistics in each histogram bin of the reconstructed angular distribution is crucial to avoid uncertainties in the final result, therefore the event selections with loose requirements on the BDT output at each value of $p_{beam}$ are preferred. However, this should be balanced against the need to suppress the large background contribution. Stricter requirements on the BDT output led to strong increases in the uncertainty of the final signal angular distribution and are therefore not preferred. The final selection criteria are chosen at each $p_{beam}$ value in order to minimize the statistical uncertainty on the determined proton form factors.

\begin{table*} [th!]
 \centering
\begin{tabular}[c]{l | c  c  c  c  c  c  c  c}
\hline
$p_{beam}$     &  $M_{inv}$     & ${|\phi^{+}-\phi^{-}|}_{lab}$  & ${(\theta^{+}+\theta^{-})}_{CM}$ & BDT & $\epsilon_{tot}$ & $\epsilon_{B}$ & S-B     \\

[GeV/$c$]      &[GeV/${c}^{2}]$ &           $[DEG]$              &          $[DEG]$                 &     &                  &   $ [10^{-6}]$ &   ratio    \\
\hline 
1.5 & $]2.1 ; 2.4[$ & $]175.0 ; 185.0[$ & $]179.65 ; 185.0[$ & > 0.314 &  0.315 & 12.2 & 1:8  \\  
1.7 & $]2.2 ; 2.5[$ & $]175.0 ; 185.0[$ & $]179.65 ; 185.0[$ & > 0.335 &  0.274 & 11.2  & 1:10 \\ 
2.5 & $]2.4 ; 2.8[$ & $]175.0 ; 185.0[$ & $]179.65 ; 185.0[$ &  > 0.280 & 0.334 &  17.5 & 1:13 \\ 
3.3 & $]2.6 ; 3.1[$ & $]175.0 ; 185.0[$ & $]179.65 ; 185.0[$ & > 0.320 & 0.295 & 13.0 & 1:5 \\
\hline
\end{tabular} 
\caption{Criteria used to select the signal ($\mu^+ \mu^-$) and suppress the background ($\pi^+\pi^-$) events for each $p_{beam}$ value. The criteria are chosen in order to keep enough signal events in each bin of the reconstructed angular distribution histogram and at the same time to suppress as many background events as possible.
The last columns list the values of the signal efficiency, background efficiency and signal-to-background ratio (S-B ratio).}
\label{tab:selection}
\end{table*}

\subsection{Angular distribution of the signal efficiency}

  \begin{figure*}[!htbp]
\centering
\begin{subfigure}{.5\textwidth}
  \centering
  \includegraphics[width=0.9\textwidth]{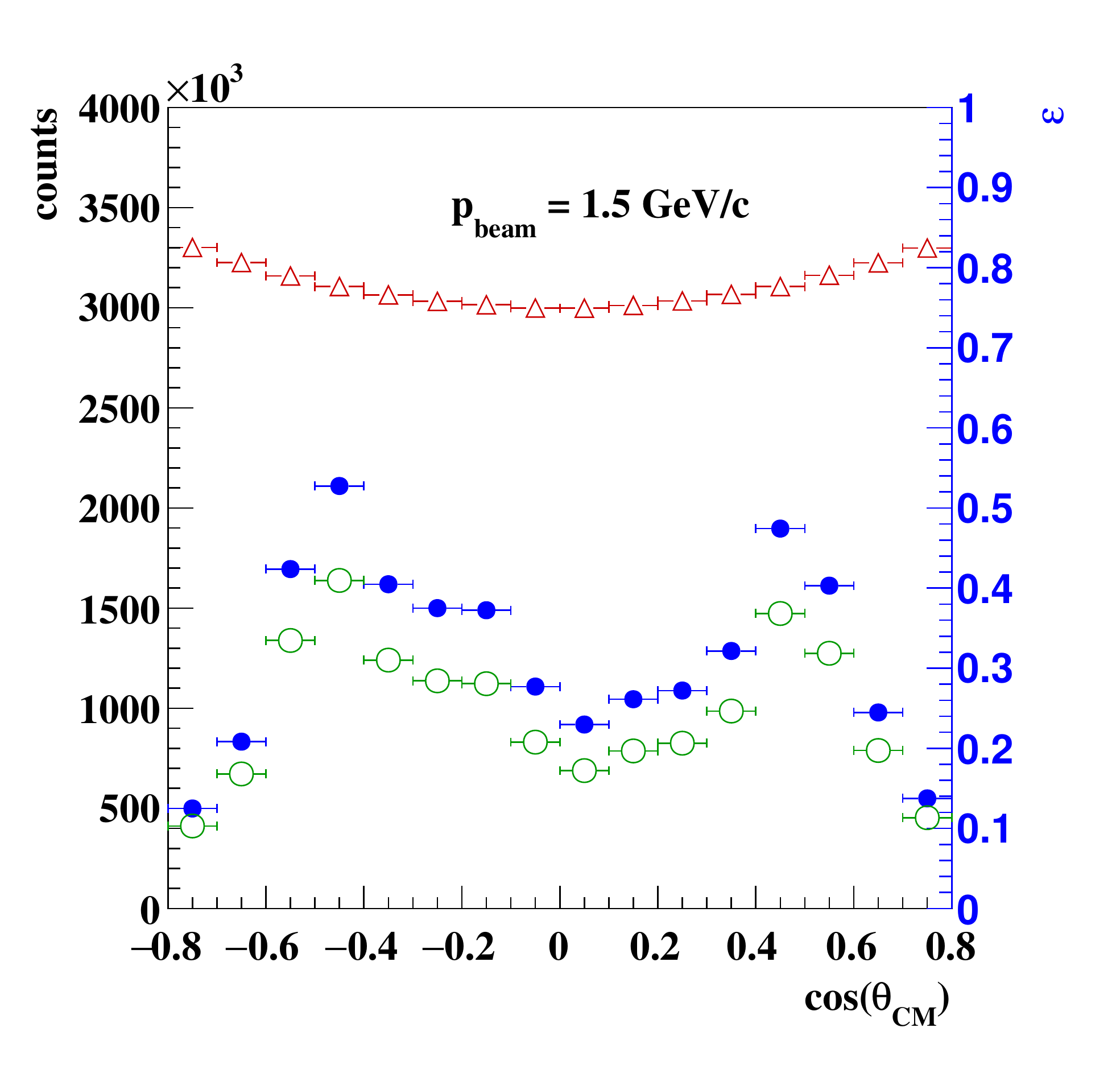}
    \caption{}
  \label{fig:sig_1_5}
\end{subfigure}%
\begin{subfigure}{.5\textwidth}
  \centering
  \includegraphics[width=0.9\linewidth]{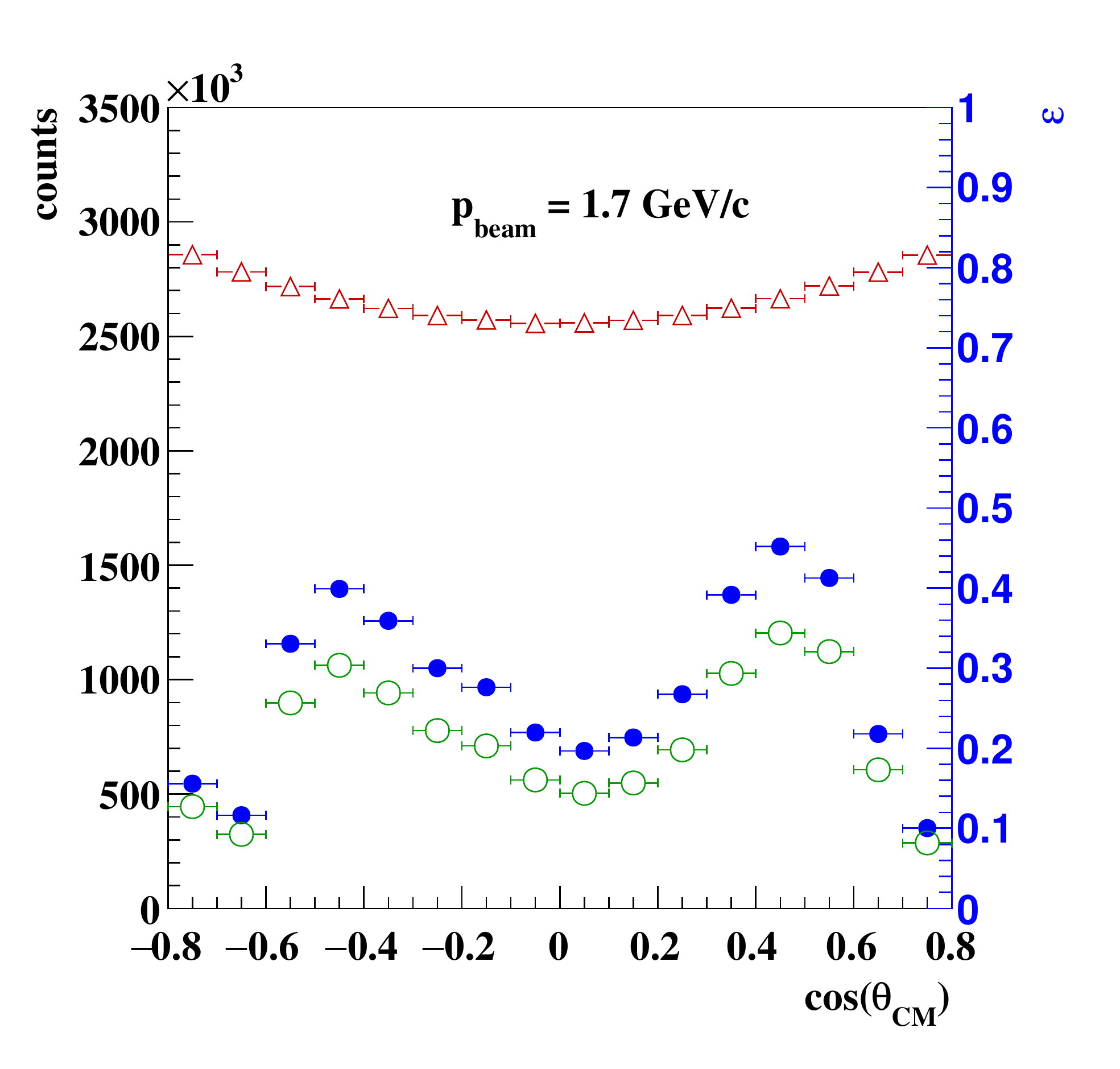}
     \caption{}
  \label{fig:sig_1_7}
\end{subfigure} \\
\begin{subfigure}{.5\textwidth}
    \centering
    \includegraphics[width=0.9\textwidth]{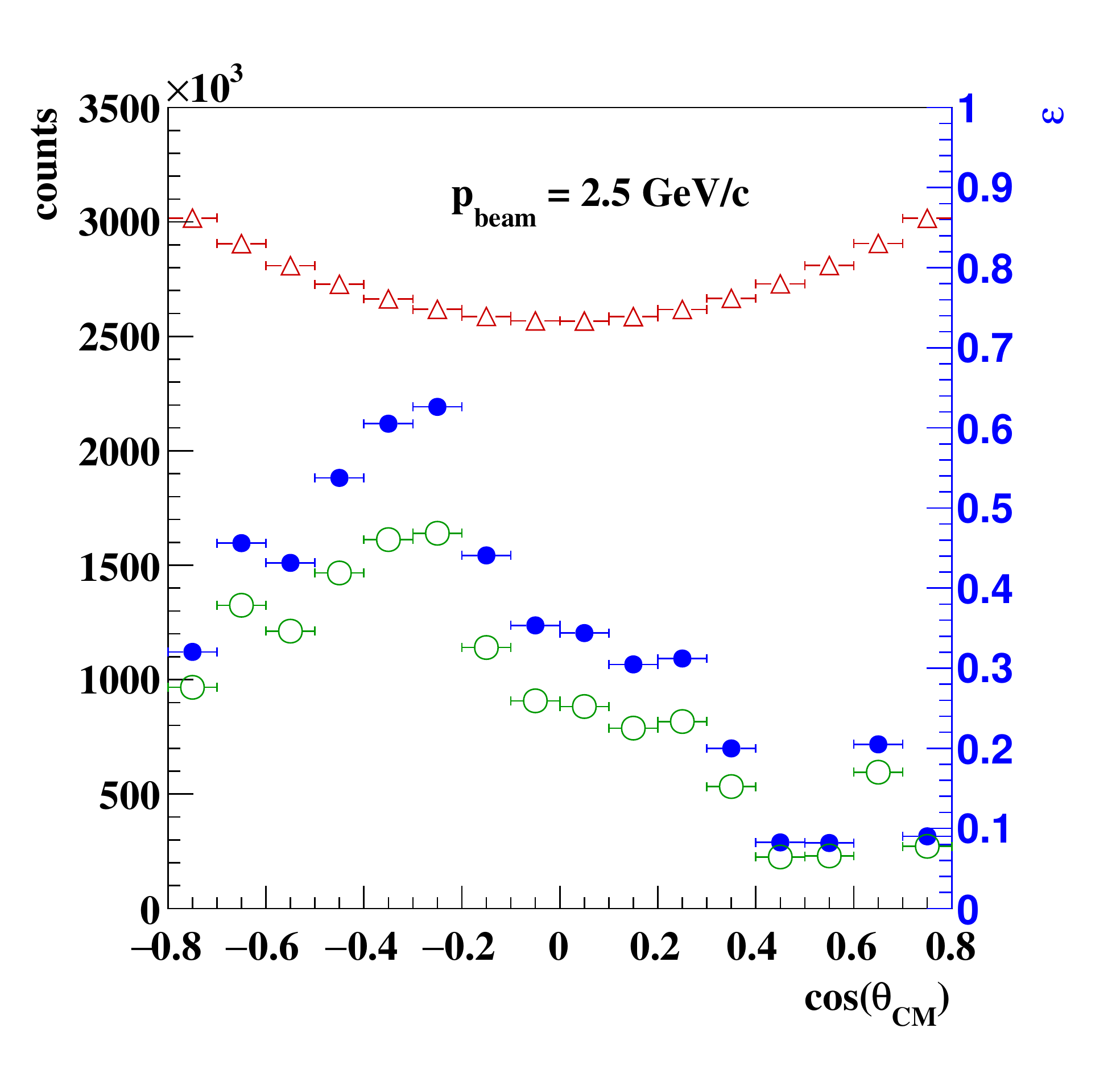}
    \caption{}
  \label{fig:sig_2_5}
\end{subfigure}%
\begin{subfigure}{.5\textwidth}
  \centering
  \includegraphics[width=0.9\linewidth]{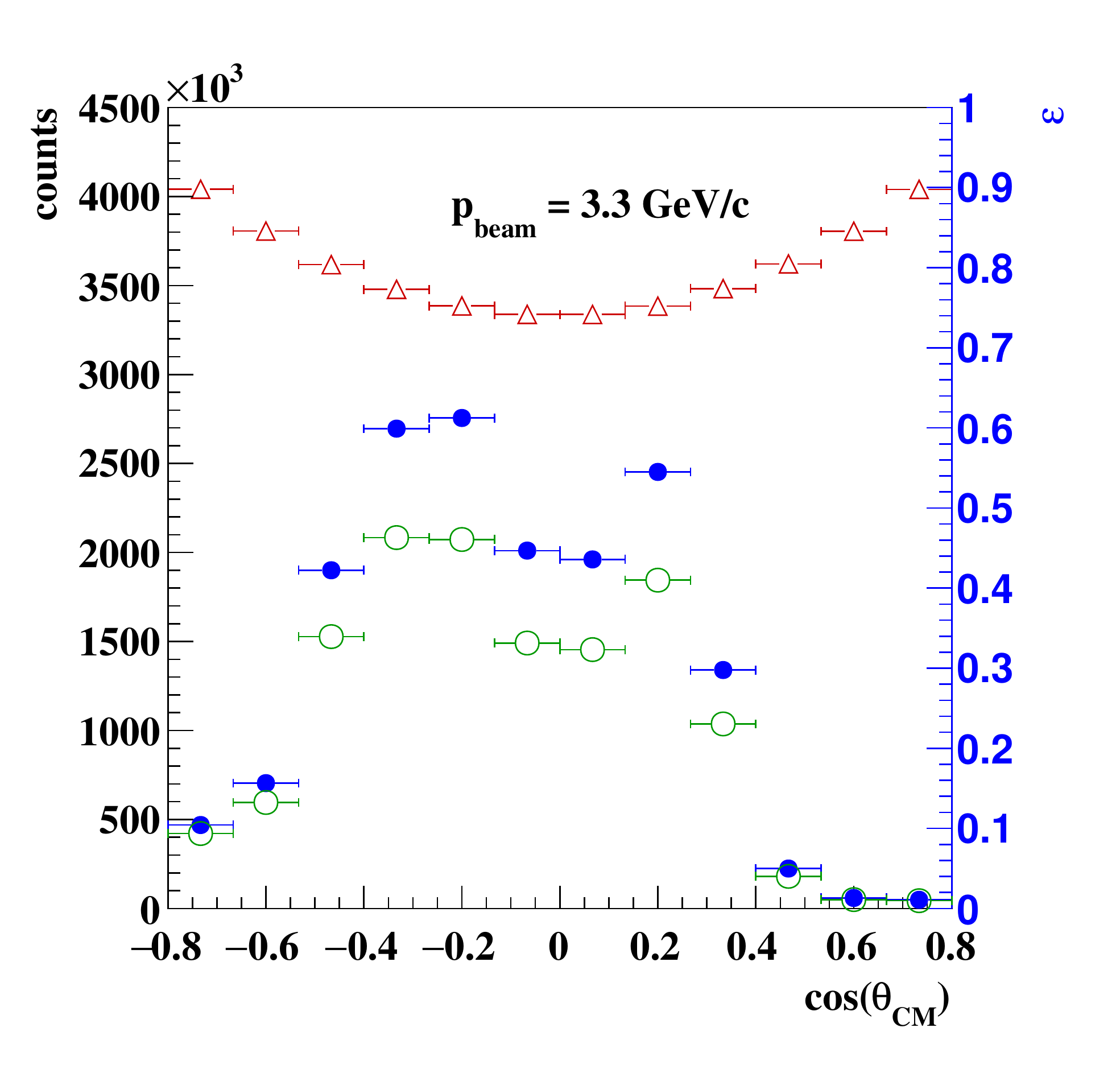}
     \caption{}
  \label{fig:sig_3_3}
  \end{subfigure}
\caption{Angular dependance of the signal efficiency $\epsilon$ (blue dots, scale on the right side) after applying the event selection conditions (Tab.\ref{tab:selection}), the MC generated signal (red open up triangles) and the selected signal events (green open circles) from sample S1 at antiproton momentum of (a) 1.5 GeV/$c$, (b) 1.7 GeV/$c$, (c) 2.5 GeV/$c$, (d) 3.3 GeV/$c$.  }
\label{fig:sigeff1}  
\end{figure*}

Figure \ref{fig:sigeff1} shows the angular dependance of the signal efficiency ($\epsilon$) (blue dots), the MC generated signal events (red open up triangles) and the selected signal events (green open circles) from sample S1 at different beam momenta. The shape of the signal efficiency distribution is determined by the selections on the BDT response. Especially at $p\bar{p}$ = 3.3 GeV/$c$, the signal efficiency drops strongly in the $0.4<$cos($\theta_{CM}$)$<0.8$ polar angle range. This is caused by the high values of the pion differential cross section in this range, which overwhelms the ability to separate signal from background. Therefore, these histogram bins will be excluded in further steps.

\subsection{Background contamination from $\bar{p}p \rightarrow \pi^+\pi^-$}

From the obtained background suppression factors, it follows that a high pion contamination, including muons from pion decay, will be expected in the $\mu$-selected data. In order to correct for it, a background subtraction procedure will be applied to the experimental data, which will introduce additional statistical uncertainty. The influence of this procedure on the precision of the extracted FF values needs to be considered in this feasibility study. In the  experiment, the measured  pion contamination  of the $\bar{p}p \rightarrow \pi^+\pi^-$ background reaction and the pion contamination  in the $\mu$-selected signal data will not exhibit identical statistical fluctuations due to the different procedures used to extract them. Therefore, two statistically independent angular distributions of the pion contamination are required to assess the background subtraction performance in this study.

A background suppression factor ($\epsilon_{B}$) of the order of $10^{-5}$ is typically achieved (see Tab. \ref{tab:selection}). The  expected number of  produced  $\bar{p}p \rightarrow \pi^+\pi^-$ background events is on the order of $10^{9}-10^{11}$ (exact numbers are listed in Tab. \ref{tab:crosssections}) assuming a time-integrated luminosity of 2 fb$^{-1}$. From that, the expected numbers of  background events after $\mu$-selection are calculated.  They include also the events with muons in the final state such as $\pi^{-}\mu^{+}$, $\mu^{-}\pi^{+}$ and $\mu^{-}\mu^{+}$, in which one or both the pions decay.

\begin{figure*}[!htbp]
\centering
\begin{subfigure}{.5\textwidth}
  \centering
  \includegraphics[width=0.9\textwidth]{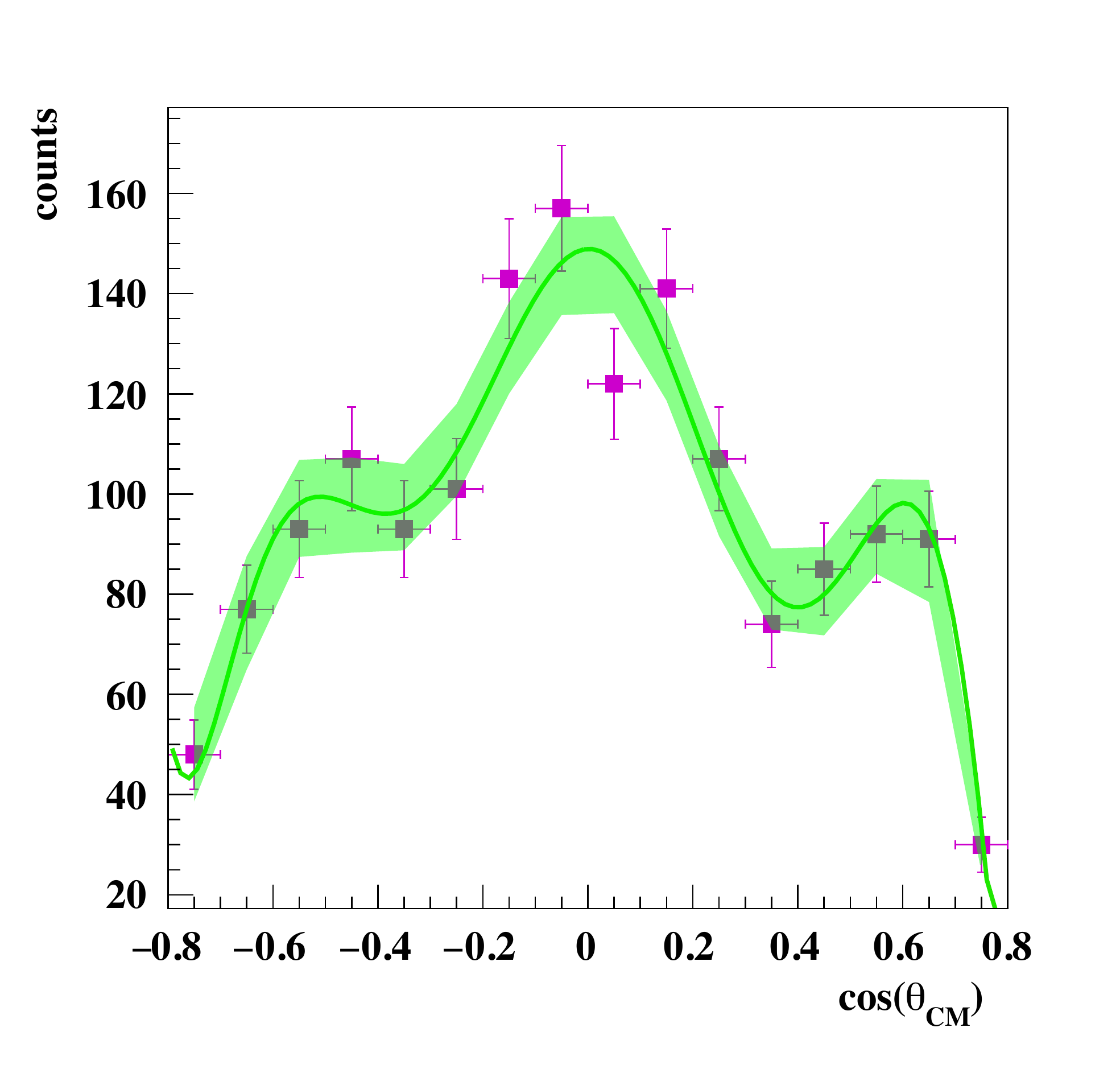}
     \caption{}
  \label{fig:Thetasm_aftercuts}
\end{subfigure}%
\begin{subfigure}{.5\textwidth}
  \centering
  \includegraphics[width=0.9\linewidth]{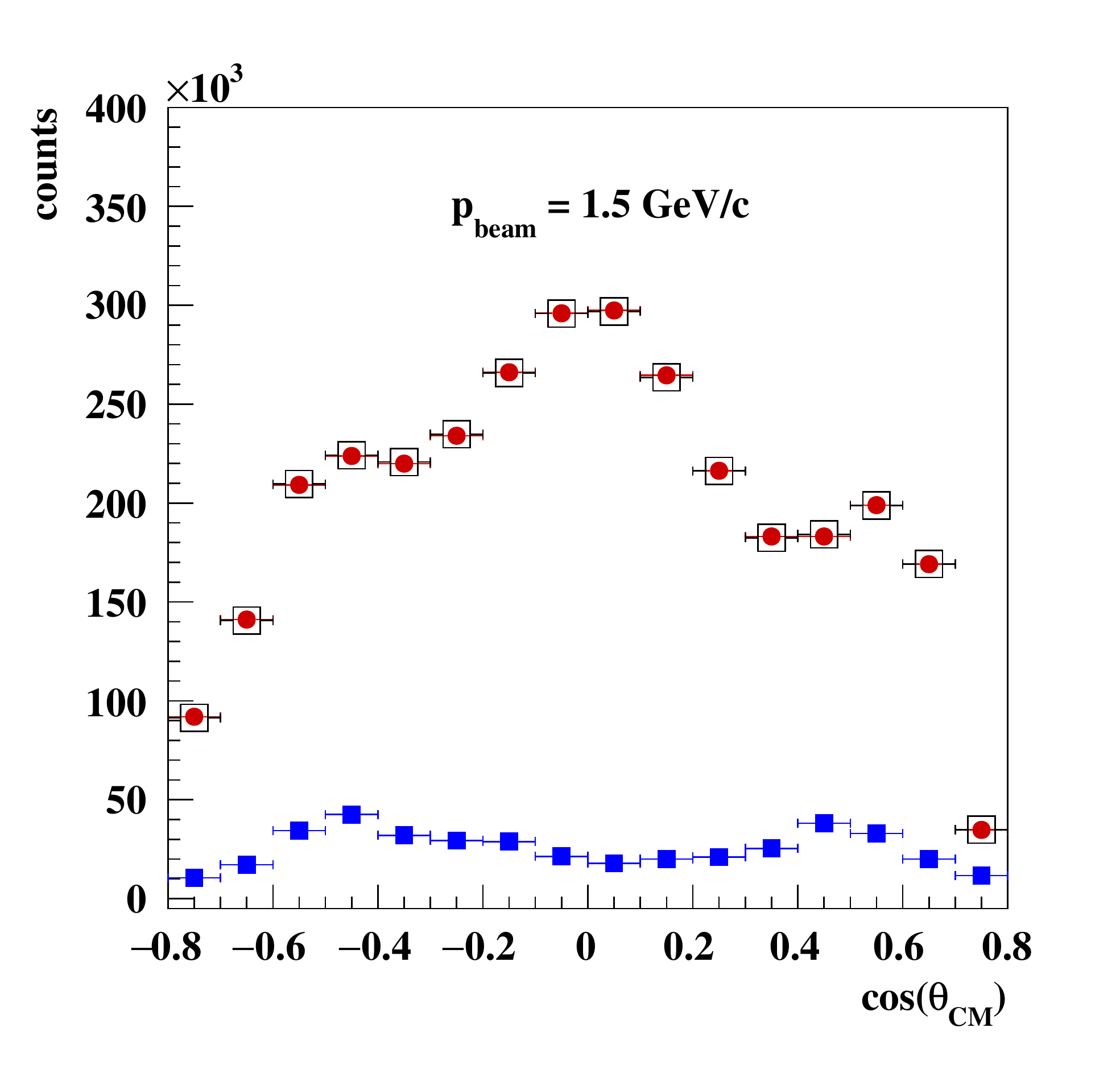}
     \caption{}
  \label{fig:newmethod}
  \end{subfigure} 
\caption{(a) Source histogram obtained with the adapted fit function $f_1$ within $\pm$1$\sigma$ error band. A reduced $\chi^2$ for the fit of $\chi^2/ndf$ = 1.59 is obtained. The function $f_1$ is used for the generation of the two statistically independent target histograms, shown in Fig.~(b)  (black open squares and red dots).  Fig.~(b) also shows the signal distribution after $\mu$-selection (blue squares).}
\label{fig:MethodII}  
\end{figure*}

The angular distribution of the pion contamination must contain the expected statistics in each bin. In this method, the $\mu$-selection is applied to the $\bar{p}p\rightarrow \pi^+ \pi^-$ reconstructed background sample and the obtained angular distribution acts as a \textit{source histogram}, which  contains a few thousand entries. The source histogram is fitted by  a polynomial of the form
\begin{equation}
f(x) =  \sum_{n=0}^{n_{max}} a_n x^n,
\end{equation}
  where the  maximum order is chosen in order to achieve an optimal value of reduced $\chi^2$. For beam momenta of 1.5 GeV/$c$ and 1.7 GeV/$c$,  \mbox{n$_{max}$ = 8} is chosen. At the largest beam momentum values, 2.5 GeV/$c$ and 3.3 GeV/$c$, a fit function with \mbox{n = 9} is optimal. This function serves as an input for a random number generator, which is used to fill a new histogram (\textit{target histogram}). The integral of the target histogram corresponds to the expected statistics. Two target histograms are created starting from different seeds. The first target histogram corresponds to the pion contamination in the selected data sample, and the second one is used for background subtraction. The obtained angular distribution in the target histograms not only contains the expected statistics, but also possess the most realistic shape, and a possible systematic uncertainty due to limited MC statistics in the background subtraction can be neglected. 
  An example of the fit function $f_1$ is shown in Fig.~\ref{fig:Thetasm_aftercuts} for the case of $p_{beam}=1.5$ GeV/$c$. Figure ~\ref{fig:newmethod} shows the obtained target histograms, which contains both the angular distribution of the pion contamination, together with the signal distribution after $\mu$-selection.

\subsection{Suppression of other relevant background channels}

Due to the high momentum and spatial resolution as well as the nearly $4\pi$ acceptance of the \PANDA detector, it will be possible to very efficiently suppress reactions of the type $\bar{p}p \rightarrow n\pi^+ n\pi^- m{\pi}^0$ with n $\geq$ 2 and m $\geq$ 0,  $\bar{p}p \rightarrow \pi^+\pi^-\omega$,  $\bar{p}p \rightarrow \pi^+\pi^-{\rho}^0$ ($\omega$ and $\rho^0$ decay into pions at a rate of nearly $100\%$).  This can be done by counting the detected charged particles in the final states and utilizing kinematical cuts.

In the  beam momentum range considered for this work, the total cross section  of the $\bar{p}p \rightarrow \pi^{+} \pi^{-} \pi^{0}$  reaction  is about seven orders of magnitude larger than the signal \cite{Boucher:2011,AlaasThesis}. In order to reach a signal contamination < 1$\%$ from this channel, a rejection factor of the order of 10$^{-9}$ must be achieved. Compared to the channel with $\pi^+\pi^-$ final state, the invariant mass of the $\pi^+\pi^-$ system in the $\pi^{+} \pi^{-} \pi^{0}$ final state is expected to be shifted drastically to smaller values and broadened because of the additional $\pi^0$. Therefore, one gains an additional rejection factor of at least of $10^{-1}$. The $\bar{p}p \rightarrow K^{+} K^{-} \pi^{0}$ reaction can be easily identified for the same reason  and also due to the high rest mass of the kaon compared to the muon mass.

\subsubsection{$\bar{p}p \rightarrow K^{+} K^{-}$ background}

Kaons from $\bar{p}p \rightarrow K^{+} K^{-}$ constitute a strong background source as well, whose cross section is of the same magnitude as the $\bar{p}p \rightarrow \pi^{+} \pi^{-}$ reaction. Therefore it is necessary to investigate if a strong enough suppression for this channel is possible.

\begin{figure}
    \centering
    \includegraphics[width=0.9\linewidth]{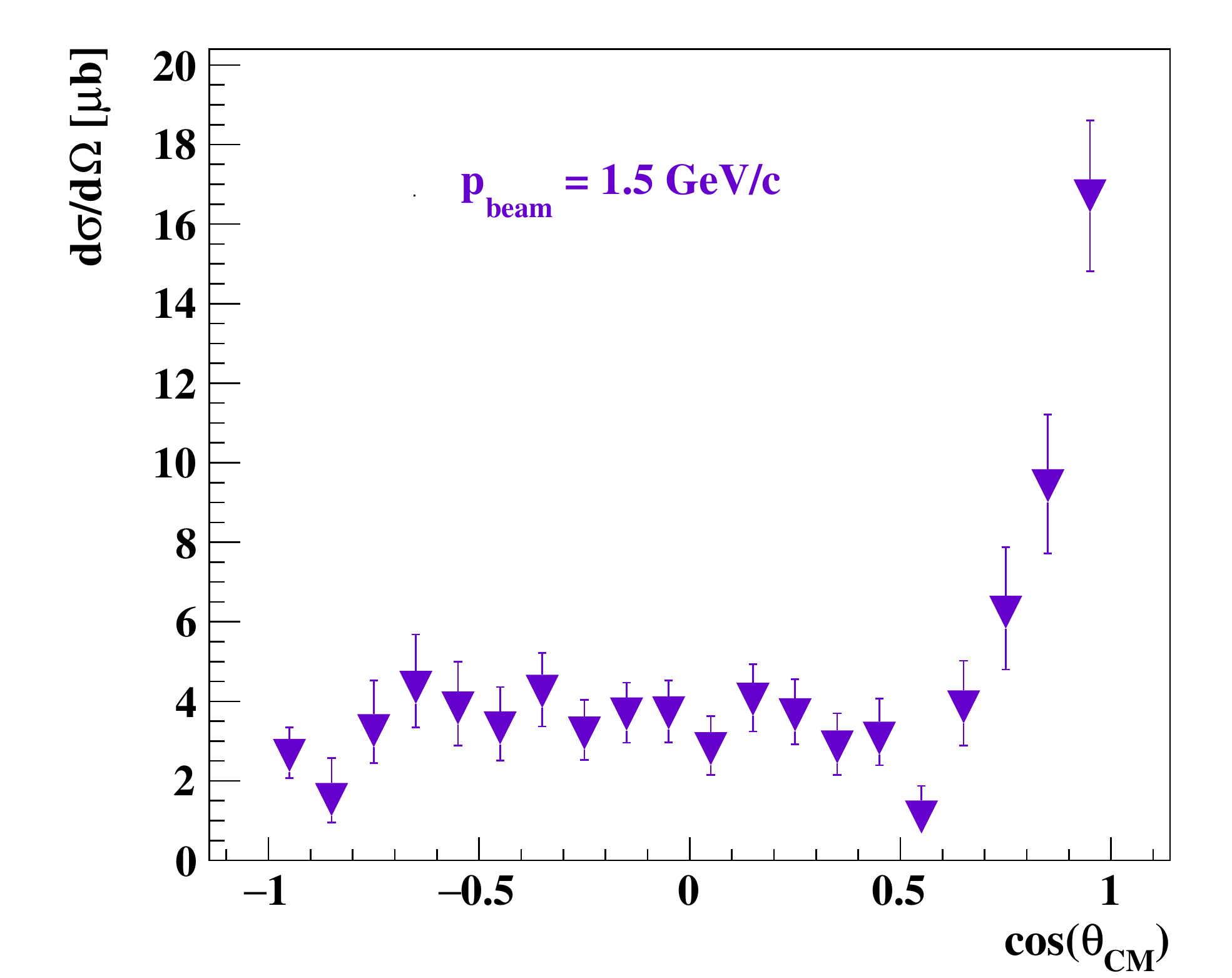} 
    \label{fig:DedxVsMom1}
    \caption{Differential cross section for the $\bar{p}p \rightarrow K^{+} K^{-}$ process as a function  of the $K^{-}$ CM emission angle from Ref. \cite{Eisenhandler:1975kx}.}
   \label{fig:kk}
\end{figure}

The differential cross section of the $\bar{p}p \rightarrow {K}^{+} {K}^{-}$ reaction was measured in 1975 by Eisenhandler et al. \cite{Eisenhandler:1975kx}. Figure \ref{fig:kk} shows the CM differential cross section as a function of cos($\theta_{CM}$)  for the negative kaon from the $\bar{p}p \rightarrow {K}^{+} {K}^{-}$ process.
For the estimation of the  rejection factor, the EvtGen generator \cite{Ryd:2005} is used to produce  phase space (PHSP) angular distributions. This estimation is assessed at the lowest value of beam momentum, where the highest precision of the FFs is achieved, as well as at the highest beam momentum, 3.3 GeV/$c$.

At 1.5 GeV/$c$, the integration over the possible angular range leads to a total cross section of 53.38 $\mu$b. Assuming a time-integrated luminosity of 2 fb$^{-1}$, this corresponds to 1.07 $\cdot$ 10$^{11}$ expected kaon events at this value of beam momentum. The total cross section of the $\bar{p}p \rightarrow K^{+} K^{-}$ channel decreases with increasing  beam momentum. In total, more than $1.05 \cdot10^8$ events are generated in the |cos($\theta_{CM}$)| < 0.8 angular range. Since the  masses of the charged kaons are  larger by a 4.7 factor than the muon rest mass, the misidentification probability for kaons is expected to be much smaller than the pion case. Hence, the kinematical cuts are much more powerful for the suppression of this process. Applying the event selection conditions reported in Tab. \ref{tab:selection}, a suppression factor better than 10$^{-8}$ is achieved for this background channel with a confidence level of 95$\%$. This corresponds to  a signal pollution < 1$\%$ with  a total signal efficiency of 31.5$\%$, therefore the contamination from this channel can be neglected. Also at $p_{beam}$ = 3.3 GeV/$c$, a signal pollution < 1$\%$ is achieved, with a total signal efficiency of 29.5$\%$.

\section{Results for feasibility at \PANDA}

This section describes the extraction of the time-like electromagnetic proton FFs, $|G_E|$ and $|G_M|$, and their ratio $R$ = $|G_E|$/$|G_M|$, from the efficiency-corrected angular distributions of the reconstructed and selected simulated data. A background subtraction is always included in these studies based on the reconstructed pion contamination distributions, as discussed in the previous section. A fit is used for the extraction of the different physical quantities and their uncertainties. At this level of the simulation, systematic uncertainties can already be estimated and included into the calculation of the total uncertainties. The proton effective FF and the total $\bar p p \to \mu^+ \mu^-$ signal cross section with their uncertainties are extracted  from the selected and efficiency corrected simulated data.

After the background subtraction, the signal efficiency correction is applied in each $i$th bin of the angular distribution of the selected simulated data:
\begin{equation}
\label{eq:effcorr}
N^{corr}_{i} = \frac{N^{reco}_{i, fluc}}{\epsilon_{i}},
\end{equation}
 with $N^{corr}_{i}$ being the efficiency corrected number of signal events, $N^{reco}_{i, fluc}$ the number of reconstructed and selected signal events after background subtraction and $\epsilon_{i}$ the signal  efficiency.
For the determination of the physical quantities, the angular distribution of efficiency-corrected signal events is fit to a function based on the differential cross section (Eq.\ref{eq:B6}):
\ba
f(\cos\theta) =  C_{1}  W_i   &\Biggl[& \frac{4{M_p}^2}{q^2} (1- {\beta_{l}}^2 \cos^2\theta) P_{1} \nn\\
&+& (1 + \frac{4{m_{\mu}}^2}{q^2}+ {\beta_{l}}^2 \cos^2\theta) P_{0} \Biggr].
\label{eqfitfunc}
\ea
The values of  $|G_E|$, $|G_M|$ and $R=|G_E|/|G_M|$ can be obtained,  from the fit parameters $P_{1}$ = $\mathcal{L} \cdot {|G_{E}|}^2$,  $P_{0}$ = $\mathcal{L} \cdot {|G_{M}|}^2$ and their ratio. Here, $\mathcal{L}$ stands for the time-integrated luminosity, for which 2~fb$^{-1}$ is assumed. $C_{1}$ is a  $q^2$-dependent constant and contains the rest masses of proton and muon:
\begin{equation}
 C_{1}  = \frac{{(\hbar c)}^2 \alpha^2 \pi}{2q^2}  \sqrt{\frac{q^2 - 4{{m}_{\mu}}^2}{q^2 - 4{{M}_{p}}^2}}. 
\end{equation}
$W_{i}$ corresponds to the width of the i-th histogram bin of the cos($\theta_{CM}$) angular distribution. 

\begin{figure*}
\centering
\begin{subfigure}{.5\textwidth}
  \centering
  \includegraphics[width=0.9\textwidth]{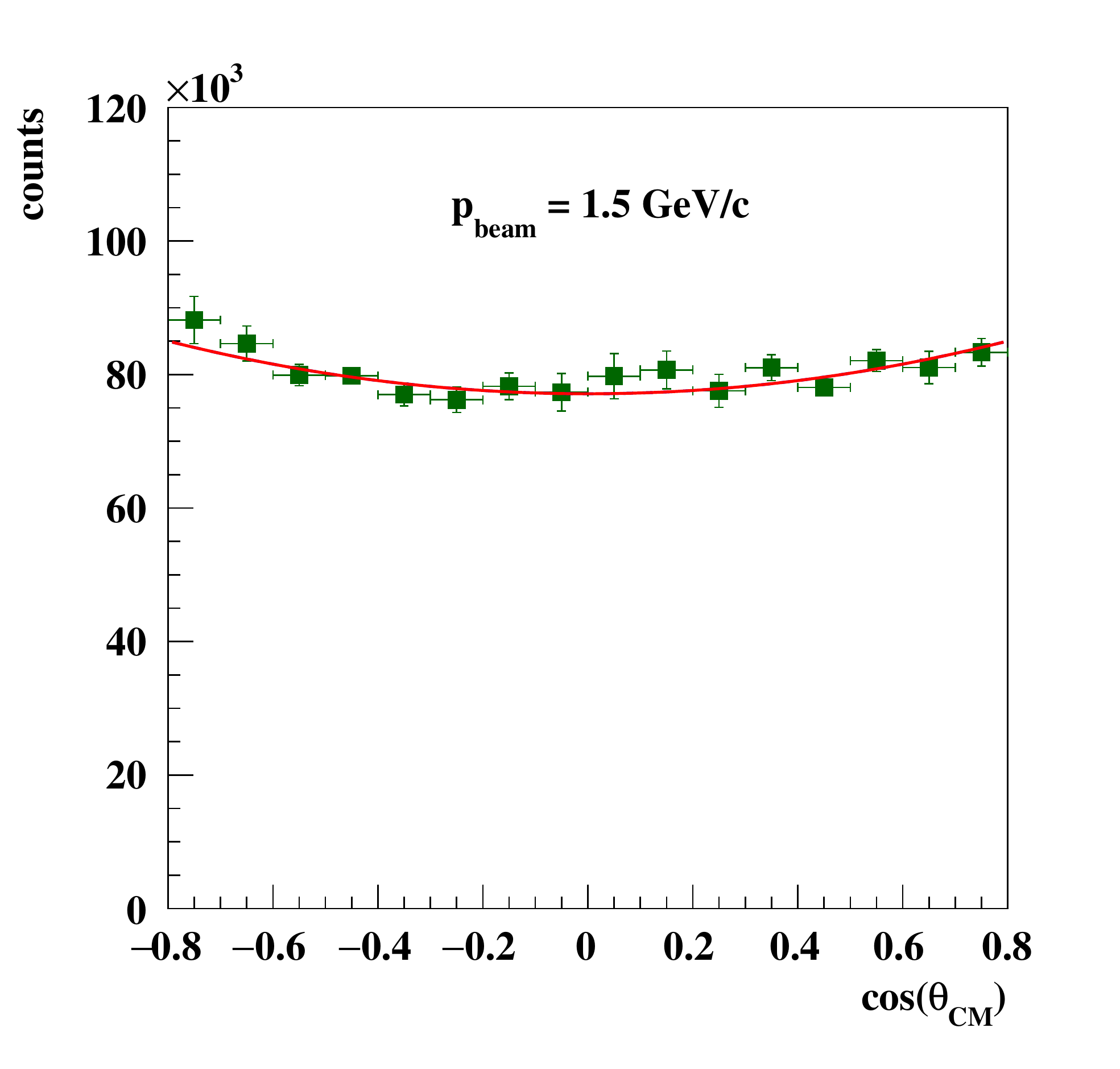}
    \caption{}
  \label{fig:FitVeryLoose_1_5}
\end{subfigure}%
\begin{subfigure}{.5\textwidth}
  \centering
  \includegraphics[width=0.9\linewidth]{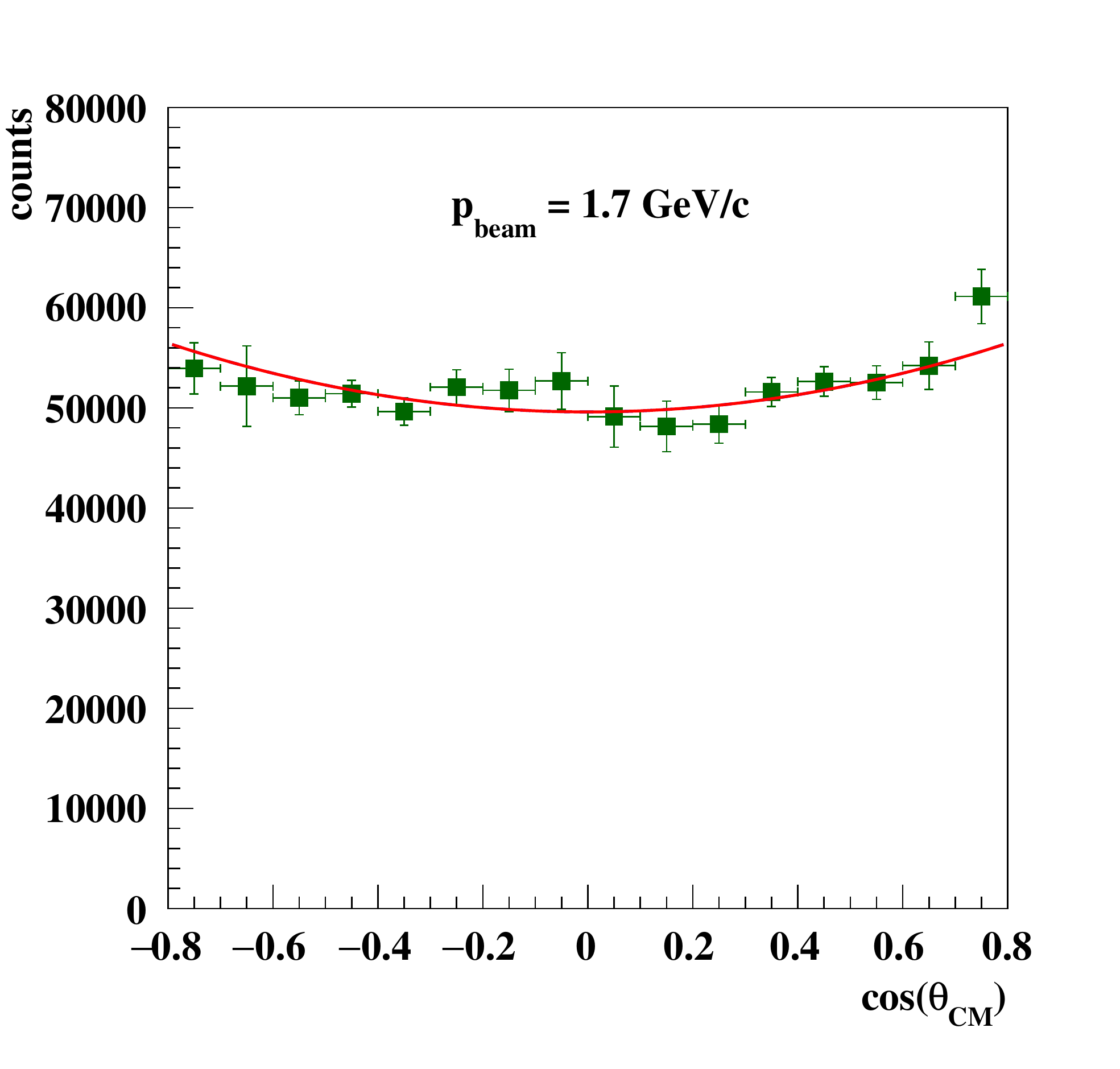}
     \caption{}
  \label{fig:FitExtremelyLoose_1_7}
  \end{subfigure} \\

\begin{subfigure}{.5\textwidth}
  \centering
  \includegraphics[width=0.9\textwidth]{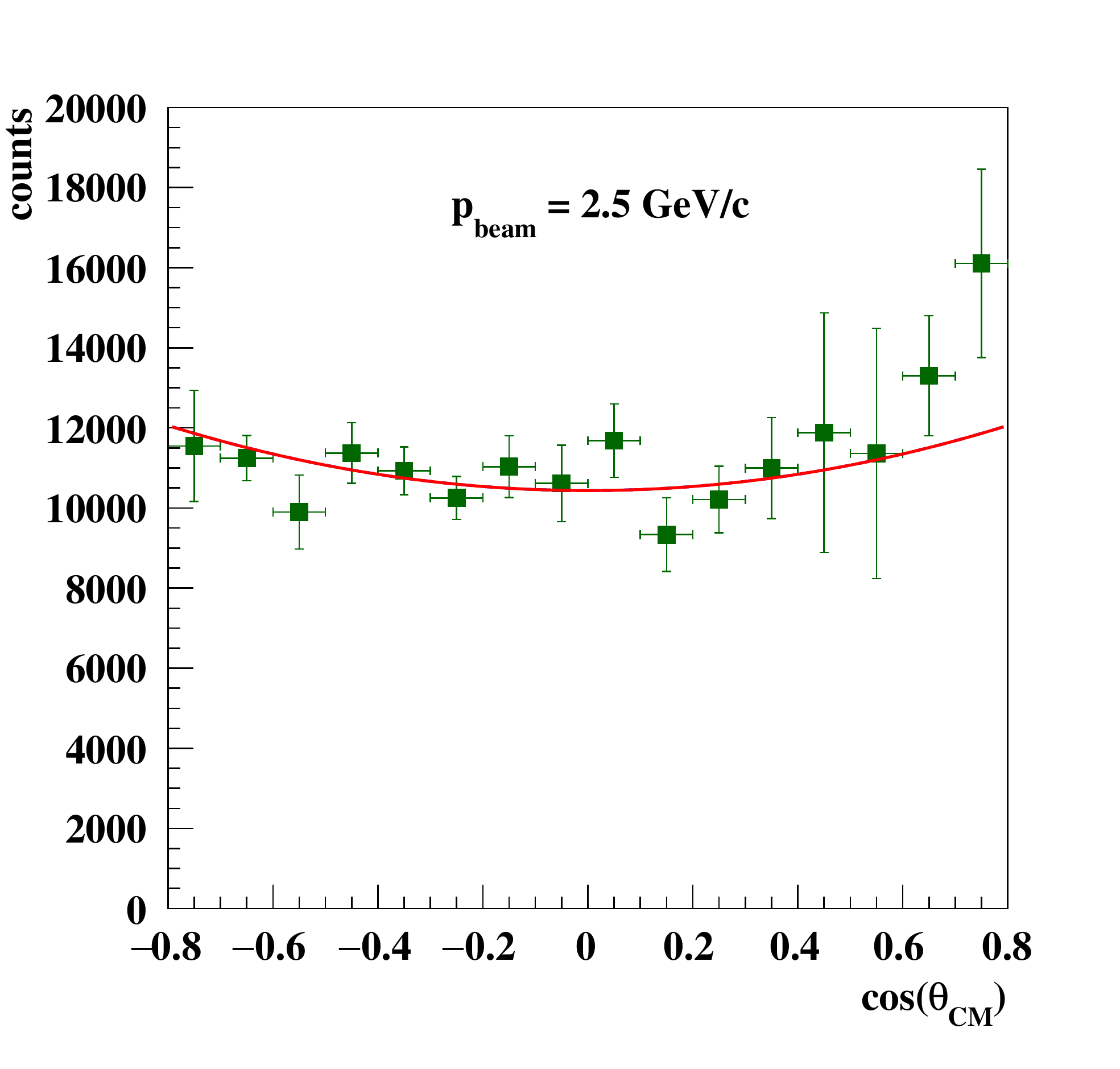}
    \caption{}
  \label{fig:FitTight_2_5}
\end{subfigure}%
\begin{subfigure}{.5\textwidth}
  \centering
  \includegraphics[width=0.9\linewidth]{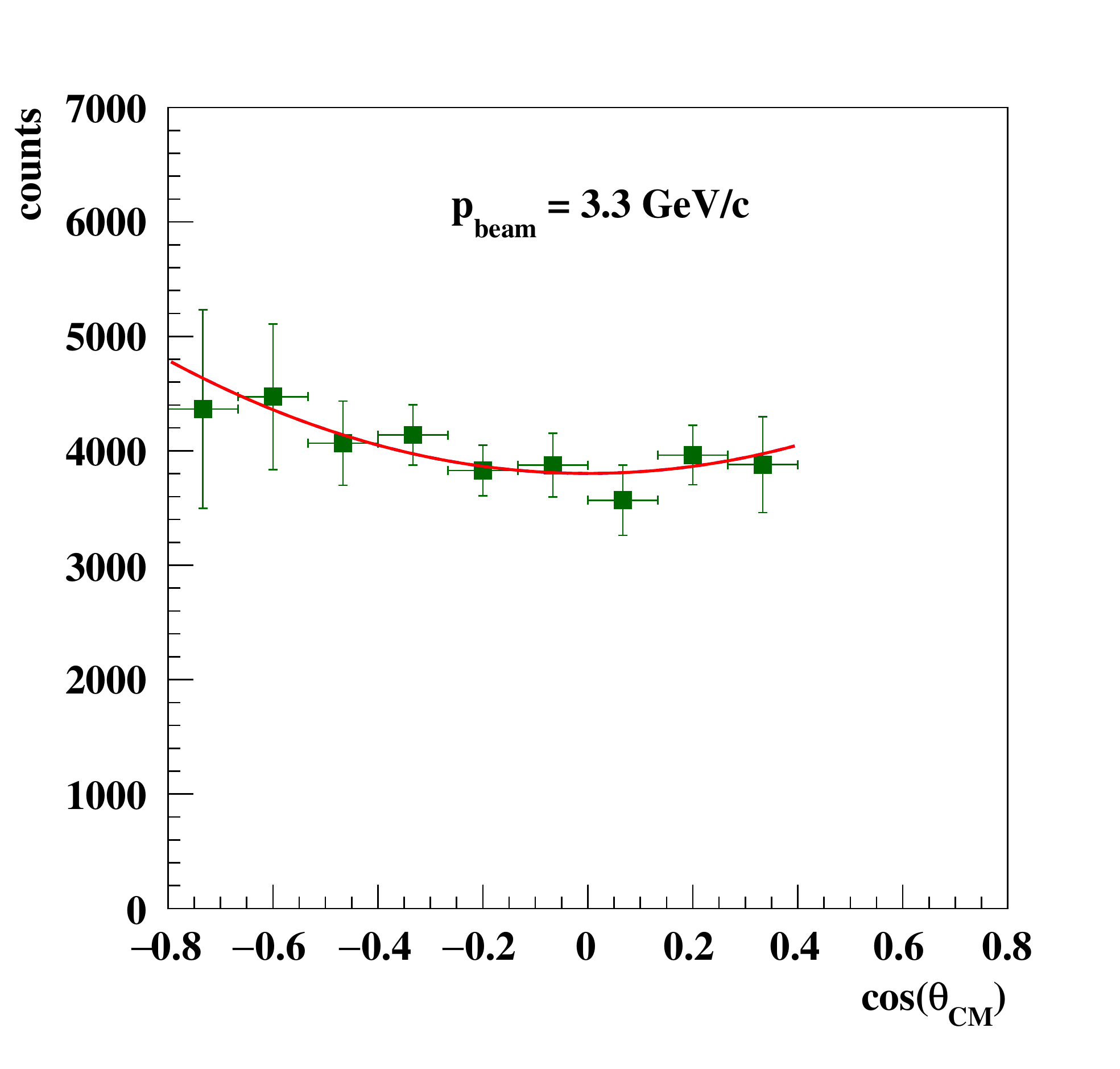} 
    \caption{}
  \label{fig:FitMiddle_3_3}
  \end{subfigure}
\caption{$cos(\theta_{CM})$ distributions for the beam momentum values indicated in each panel. The green squares with error bars represent the selected signal yields after background subtraction, corrected by the signal efficiency. The fit function (Eq.~\ref{eqfitfunc}) (red solid line) is used to extract the proton FFs and their uncertainties.}
\label{fig:MethodII:low}  
\end{figure*}

Figure \ref{fig:MethodII:low} shows the resulting angular distributions for the signal after the efficiency correction along with the corresponding fit.

\begin{table}[th!]
 \centering
  \begin{adjustbox}{max width=0.99\textwidth}
\begin{tabular}[l]{  l | c  c  c  c }  
\hline
    $p_{beam}$ [GeV/$c$]    &     $R$  & $\Delta R$ & $\frac{\Delta R}{R} [\%]$ & ${\chi}^2$/ndf \\   
          \hline      
    1.5 &  1.02 & 0.05 &   5 &   0.85 \\
    1.7 &  0.99 & 0.07 &   7 & 1.12 \\ 
    2.5 & 1.08 & 0.16  &  14 & 1.13 \\
    3.3 & 0.99 & 0.36  &  37 & 0.86 \\
  \hline
\end{tabular}
\end{adjustbox}
\caption{Extracted value and statistical precision of  $R=|G_{E}|/{|G_{M}|}$ at each considered value of beam momentum ($p_{beam}$ = 1.5, 1.7, 2.5 and 3.3 GeV/$c$). The studies are based on the assumption of $R$ = 1.}
\label{tab:fitresults1_m2}
\end{table}

\begin{table*}[th!] 
\centering
\begin{tabular}[l]{  c c   | c  c  c  c  c  c  }  
\hline
$p_{beam}$ & $|G_{M}|$ (model)       & $|G_{E}|$ & $\Delta|G_{E}|$ &  $\Delta|G_{E}|/|G_{E}|$ &$|G_{M}|$ & $\Delta|G_{M}|$ &  $\Delta|G_{M}|/|G_{M}|$ \\ 

[GeV/c]  & &  & &  $[\%]$ &  & &   $[\%]$ \\
\hline
1.5 & 0.1403 & 0.142   &  0.004        &      3.1                         &  0.139  &      0.002         & 1.5 \\
1.7 & 0.1213  &  0.121  & 0.006      &       5.1                        &    0.122   &       0.003    &     2.2     \\
2.5 &     0.0703   &    0.074    & 0.008          &     10.2       &    0.068    &      0.003        & 4.4      \\
3.3   &    0.0436  &   0.043        & 0.012        &    26.9       &   0.044   &     0.004          &     9.6   \\
\hline
\end{tabular}
\caption{Extracted value and statistical precision of  $|G_{E}|$  and ${|G_{M}|}$  at each value of beam momentum ($p_{beam}$ = 1.5, 1.7, 2.5 and 3.3 GeV/$c$).  The theoretical values of the magnetic FF, $|G_M|$ (model), which are based on the FF model for the parameterization of $|G_M|$ from Ref. \cite{TomasiGustafsson:2001za} are shown for comparison.}
\label{tab:fitresults2_m2}
\end{table*}
Tables  \ref{tab:fitresults1_m2} and \ref{tab:fitresults2_m2} show the extracted values of $R$, ${|G_{E}|}$  and ${|G_{M}|}$ with their uncertainties at each value of beam momentum ($p_{beam}$ = 1.5, 1.7, 2.5 and 3.3 GeV/$c$). The results are consistent with the corresponding theoretical values used as input to the simulations within one sigma. The 4${M_{p}}^2/q^2$ factor in the differential cross section formula suppresses the $|G_E|$ term, so that at larger values of $q^2$ the cross section is dominated by $|G_M|$, which also leads to a larger uncertainty in the measured values of $|G_E|$.

\subsection{Integrated cross section and the effective proton FF}
The integrated cross section of the $\bar{p}p \rightarrow \mu^+\mu^-$  process is calculated for each value of $q^2$ as
\begin{equation}
    {\sigma}_{int} = N^{corr}/\mathcal{L},
 \end{equation}
with $\mathcal{L}$ = 2fb$^{-1}$. Table \ref{tab:results_sigma_m2} shows the obtained values of the integrated cross section at each considered value of beam momentum.

\begin{table*}[ht!]
  \centering
  \begin{adjustbox}{max width=\textwidth}
\begin{tabular}[l]{ c | c |  c | c c}  
\hline
  							         $p_{beam}$                   &          $q^2$       &    $\sigma$   (theoretical value)    &           $\sigma \pm \Delta \sigma$    (extracted)    &     $\Delta \sigma/\sigma$    (extracted)     \\
							$[$GeV/$c]$ 	   &     $ [$(GeV/$c$)$^2]$	 &     	  [pb]  &	[pb]	&      $[\%]$         \\  
\hline 
							                1.5				&  5.08 	&	640.7		&		640.6    $\pm$	4.2    	& 0.6   \\
								  	1.7				& 5.40	&      414.9		&		413.9     $\pm$	5.9	   & 1.4   \\
							          	2.5				& 6.77		& 89.19		&		91.48      $\pm$	1.92             & 2.1  \\
							    	        3.3				&  8.20		& 24.83		&		 24.91     $\pm$	0.69	   & 2.8  \\
\hline 
\end{tabular}
\end{adjustbox}
\caption{Extracted statistical precisions of the integrated cross section of the $\bar{p}p \rightarrow \mu^+\mu^-$ signal process together with the calculated values based on Eq.~\ref{eq:B6}, in -0.8 < cos($\theta_{CM}$) < 0.8 angular range.}
\label{tab:results_sigma_m2}
\end{table*}
From the results one can conclude that the integrated cross section of the signal process can be determined with high accuracy at \PANDA.  The  proton effective  FF can be determined from the integrated cross section in the  |cos($\theta_{CM}$)| < $\bar{a}$ range ($\bar{a}$ = 0.8) using
\begin{equation}
\label{eq:effektiveFF}
|F_{p}|   = \sqrt{\frac{\sigma_{int}(q^2)}{\frac{\pi{\alpha}^{2}}{2q^2} \frac{\beta_{\ell}}{\beta_{p}}  \bigg[ (2 - {\beta}_{\ell}^{2}) + \frac{1}{\tau} \bigg ] \;  \bigg[2 \bar{a} +  \frac{2}{3} A_0 \; \bar{a}^3 \bigg]} },
\end{equation}
with 
\begin{equation*}
A_{0}={\beta}_{\ell}^{2} \; \frac{1-{\frac{1}{\tau}}}{2-{{\beta}_{\ell}}^{2}+\frac{1}{\tau}},
\label{eq:factors_effFF}
\end{equation*}

being $\tau = q^2/4M^2_p$ , $\beta_{\ell,p} =\sqrt{1-4M^2_{\ell,p}/q^2}$.

\begin{table*}[th!]
 \centering
  \begin{adjustbox}{max width=\textwidth}
\begin{tabular}[l]{ c | c | c | c  c}  
\hline
  							   $p_{beam}$                   &          $q^2$           &   $|F_p|$ (model) &       $|F_p|$  $\pm$ $\Delta |F_p|$  (extracted) &     $\Delta |F_p|/|F_p|$  (extracted)  \\   	
										$[$GeV/$c]$ 	   &     $ [$(GeV/$c$)$^2]$	 &     			&                   &       $[\%]$         \\  
\hline 
							                1.5				&  5.08 		&  0.1403			&		0.1402     $\pm$ 	0.0005    	& 0.3   \\
								  	1.7				& 5.40		& 0.1213			&		0.1210     $\pm$ 	0.0009	       & 0.7   \\
							          	2.5				& 6.77		& 0.0703		       &		0.0712       $\pm$ 	0.0007            & 1.1	  \\
							    	        3.3				&  8.20		& 0.0436		       &		 0.0437      $\pm$ 	 0.0006	   & 1.4  \\
\hline 
\end{tabular}
\end{adjustbox}
\caption{Extracted values and statistical precisions of the effective proton FF, ${|F_{p}|}$. The third column is the theoretical value (simulation input).}
\label{tab:results_effFF_m2}
\end{table*}
The extracted relative statistical uncertainty of the effective FF (Tab. \ref{tab:results_effFF_m2}) ranges between 0.33$\%$ and 1.39$\%$ for beam momenta between 1.5 and 3.3 GeV/$c$. As a systematic uncertainty, the contribution from the luminosity measurement can be calculated as $\Delta |F_p|/ |F_p|$ (syst.) = $\pm$~2$\%$, assuming a relative uncertainty of the luminosity of 4$\%$ at all values of $q^2$.

\subsection{Systematic uncertainties}

Since only MC simulated data are currently available, naturally a precise estimation of the systematic uncertainties is not possible at the present time. However, several sources of systematic uncertainties can already be estimated based on the MC study and will be discussed in the following.

\subsubsection{Luminosity measurement}

\PANDA will determine the luminosity $\mathcal{L}$ exploiting the well known  elastic $\bar{p}p$ scattering. $\mathcal{L}$ will be measured with a relative systematic uncertainty from 2.0$\%$ to 5.0$\%$, depending on the beam momentum, the knowledge of the differential cross section parameters and the $\bar{p}p$ inelastic background contamination \cite{LumiTechRep}. In this estimation, a relative uncertainty of $\Delta\mathcal{L}/\mathcal{L}$ = 4.0$\%$ is assumed at all beam momenta. This corresponds to a relative uncertainty on the determination of the proton FFs of 2.0$\%$.

\subsubsection{Choice of event selections}

The signal and background efficiencies are mainly affected by the selections on the BDT outputs. The contribution from the choice of event selections to the total systematic uncertainty is determined at each beam momentum by varying the value of the BDT output selection around the reference value. The spread of the values of the proton FFs using different BDT selections is taken as the systematic uncertainty due to this source.

\subsubsection{Choice of histogram binning}

 The $\cos(\theta_{CM})$ distributions binning has an effect on the values of the extracted quantities and their uncertainties. In order to compare to the results determined for the $\bar{p}p \rightarrow e^+e^-$ reaction from Ref. \cite{DmitryAlaaPaper}, the same binning (16 bins) is chosen at beam momenta $p_{beam}$ = 1.5, 1.7 and 2.5 GeV/$c$. At $p_{beam}$ = 3.3 GeV/$c$, wider bins are chosen (12 bins)  since the data points show larger statistical fluctuations. The difference between the results obtained with the different binning is calculated at 3.3 GeV/$c$ and is used for the determination of the uncertainty due to this source.

\subsubsection{Asymmetry contributions to cos($\theta_{CM}$)}

In this work, no radiative corrections are included, since no calculations for the muon channel exist. A symmetric angular distribution in cos($\theta_{CM}$) is assumed in this work as a consequence of the one-photon exchange approximation.

In Ref. \cite{DmitryAlaaPaper} two-photon exchange for the electron channel is discussed, which introduces asymmetry contributions to the angular distribution \cite{Pa:2015, Guttmann:2011b}. The contribution of the two-photon exchange to the cross section for the electron channel is expected to be negligible, being less than 1$\%$ \cite{VandeWiele:2013}. The contribution of the interference term between one- and two-photon-exchange is symmetric under interchange of electron and positron and can be removed from the angular distribution by adding both angular distributions \cite{Gakh:2005}. This strategy will be also applied for the muon channel.

\subsubsection{Pion background}

The cross section of the background channel $\bar p p \to \pi^+ \pi^-$ will be measured at \PANDA with a very
high precision due to its large cross section (see Tab.~\ref{tab:crosssections}). The same data samples will be used to extract the signal and background processes. Therefore, systematic uncertainties due to the modeling of the differential cross section of this process used in simulations or due to the detector performance are expected to be negligible.  In addition, the influence of the shape of the pion background distribution on the extracted precision of the form factors is investigated and found to be also negligible.

The  contributions to the systematic uncertainties of the proton FFs  are summarized in in  Tab. \ref{tab:results_total_m2}.

\subsection{Total relative uncertainties}

An overview of the statistical and systematic contributions to the relative total uncertainty of the FFs and the ratio $R$ is given in Tab. \ref{tab:results_total_m2}. The largest sources of systematic uncertainties are related to the choice of histogram binning, the event selections and the luminosity measurement. The total uncertainty is listed for all the considered beam momenta.
\begin{table*}[th!]
\begin{center}
\begin{tabular}{l | c | c | c  | c c c  || c}  
\hline\noalign{\smallskip}
							   &          $p_{beam}$                   &          $q^2$           &             Relative statistical     	   &   \multicolumn{3}{c||}{Relative systematic uncertainty}        &      Total     \\   
							   & 			[GeV/$c$] 	   &      [(GeV/$c$)$^2]$	 &     		uncertainty					&    {\it Binning}       &  {\it Cuts}  &        {\it  Luminosity}         &      \\  
\noalign{\smallskip}\hline\noalign{\smallskip}
							   &           1.5				&  5.08 				&		3.1$\%$	& 		  -		    &  0.1$\%$ 		& 2.0$\%$	           & 3.7$\%$  \\
$\frac{\Delta|G_{E}|}{|G_{E}|}$      &		1.7				& 5.40				&		5.1$\%$	& 		-	         	& 1.3$\%$	 & 2.0$\%$		   & 5.6$\%$  \\
							    &      	2.5				& 6.77				&		10.2$\%$	& 		-	                     &  4.2$\%$ 		& 2.0 $\%$			   & 11.2$\%$  \\
							    &	        3.3				&  8.20				&		26.9$\%$	&      	0.9$\%$   		&   0.9$\%$		 & 2.0$\%$			   & 27.0$\%$  \\

\hline
							   &           1.5				&  5.08 				&		1.5$\%$	& 		-		&  < 0.1$\%$		& 2.0$\%$			   & 2.5$\%$  \\
 $\frac{\Delta|G_{M}|}{|G_{M}|}$    &		1.7				& 5.40				&		2.2$\%$	& 		-		& 0.5$\%$ 		& 2.0$\%$			   & 3.0$\%$  \\
							    &      	2.5				& 6.77				&		4.4$\%$	& 		-	       &  0.5$\%$			 & 2.0$\%$			   & 4.9$\%$  \\
							    &	        3.3				&  8.20				&		9.6$\%$	& 		< 0.1 $\%$  & 1.4$\%$ 	& 2.0$\%$			   & 9.9$\%$  \\
\hline

							   &           1.5				&  5.08 				&		5$\%$	& 	  -    		  & 	0.1$\%$		      &		-	   	& 5$\%$  \\
$\frac{\Delta R}{R}$                        &		1.7				& 5.40				&		7$\%$	& 		-		&	 2.3$\%$			&	 -  		& 7$\%$  \\
							    &      	2.5				& 6.77				&		14$\%$	& 		-		& 	 4.7$\%$			&	  - 		& 15$\%$  \\
							    &	        3.3				&  8.20				&		37$\%$	& 		1.0$\%$    & 	3.0$\%$ 		&	-	   	& 37$\%$  \\

\noalign{\smallskip}\hline
\end{tabular}
  \end{center}
\caption{Statistical and systematic uncertainties, which contribute to the relative total uncertainty of ${|G_{E}|}$, ${|G_{M}|}$ and of the ratio $R=|G_E|/|G_M|$.}
\label{tab:results_total_m2}
\end{table*}

\begin{figure}
\centering
\includegraphics*[width=1.0\linewidth]{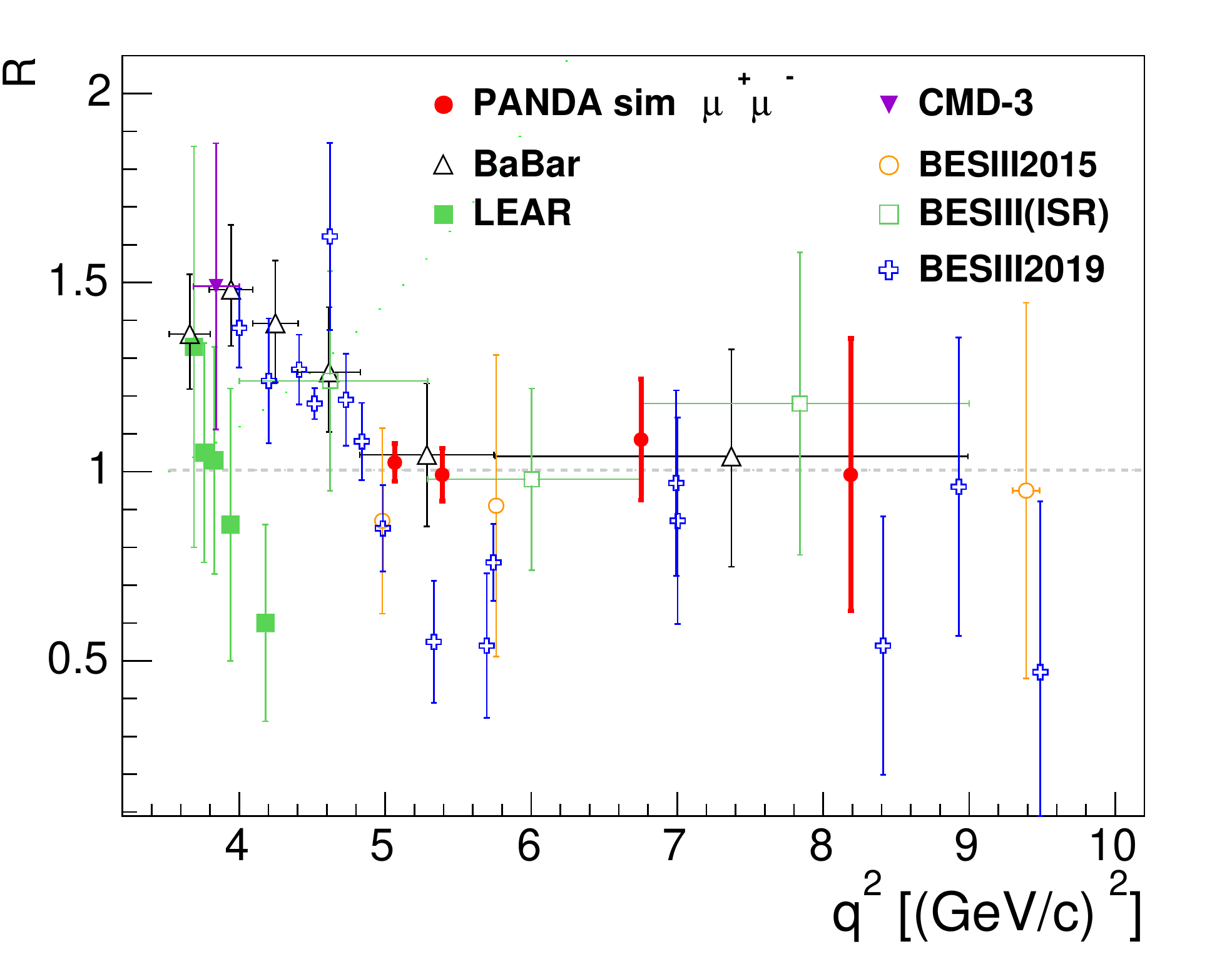}
\caption{Values, with total uncertainties, obtained for the $R = |G_{E}|/|G_{M}|$ ratio for the $\bar{p}p \rightarrow \mu^+\mu^-$ channel at different values of $q^2$ (red points). $R$ = 1 is the simulation input. Also shown are the currently existing data from Ref. \cite{Bardin:1994am} (green squares), from Ref. \cite{Lees:2013b} (open black triangles), from Ref. \cite{Ablikim:2015} (open orange  circles), from Ref. \cite{Akhmetshin:2016} (magenta down triangle), from Ref.~\cite{Ablikim:2019njl} (open green squares), and from Ref.~\cite{Ablikim:2019eau} (open blue crosses).}
\label{fig:R_final}
\end{figure}

The results show that the total relative uncertainty, $\Delta R$/$R$, ranges between 5$\%$ and 37$\%$ for  $q^2$ between 5.08 and 8.20 (GeV/c)$^2$. The estimated values of the total relative uncertainty $\frac{\Delta|G_{M}|}{|G_{M}|}$ lie between 2.5$\%$ and 10$\%$, while those for $\frac{\Delta|G_{E}|}{|G_{E}|}$ between 3.7$\%$ and 27.0$\%$. Figure \ref{fig:R_final} shows the final results obtained for $R \pm \Delta R$, including all considered statistical and systematic uncertainties. The results show that $|G_{E}|$, $|G_{M}|$ and their ratio are expected to be measured with high precision at \PANDA. At lower beam momenta, the statistical precision increases due to the increasing cross section of the signal reaction; therefore, the highest precision of the time-like proton FFs will be obtainable at the lowest possible value of $q^2$ = 5.1 (GeV/$c$)$^2$.

\section{Test of lepton universality} \label{sec:phase3}

Since the lepton universality is a fundamental feature part of the Standard Model, a violation of this universality would be a sign for new physics beyond the Standard Model. The only hints for the violation of lepton universality exist so far from experiments such as BaBar, Belle and recently LHCb (CERN) \cite{Aaij:2014}. The LHCb experiment measured the ratio of the branching fractions of the $B^+ \rightarrow K^+\mu^+\mu^-$ and $B^+ \rightarrow K^+e^+e^-$ decays using proton-proton collision data. The ratio of the branching fractions, denoted as $R_K$, within a fixed range of the di-lepton mass squared from $q^2_{min}$ to $q^2_{max}$ is given by
\begin{equation}
R_K [q^2_{min}, q^2_{max}] = \frac{\int^{q^2_{max}}_{q^2_{min}} dq^2 \frac{d\Gamma(B^+\rightarrow K^+ \mu^+\mu^-)}{dq^2}}{\int^{q^2_{max}}_{q^2_{min}}dq^2 \frac{d\Gamma(B^+\rightarrow K^+ e^+e^-)}{dq^2}},
\end{equation}
where $\Gamma$ stands for the $q^2$-dependent partial width of the B meson decay. Details of the measurement can be found at Ref. \cite{Aaij:2014}.
\\\\
A calculation of the Standard Model prediction for $R_{K}$ predicted a value of unity within an uncertainty of $\mathcal{O}$($10^{-3}$) by Ref. \cite{Bo:2007, Bouchard:2013}. More recent calculations, which have been performed by \cite{Bordone:2016} showed that the largest theoretical uncertainty of $R_K$ is due to QED corrections, and result in a relative uncertainty of $\approx$ 1-2$\%$. 
\\\\
In the LHCb measurement, a time-integrated luminosity of 3 fb$^{-1}$ was achieved at center of mass energies between 7 and 8 TeV. The measurement was performed in the  1 < $q^2$ < 6 (GeV/c)$^2$ range, where $q^2$ corresponds to the  di-lepton invariant mass squared. The ratio of branching fractions was  
\begin{equation}
R_K = 0.745^{+0.09}_{-0.074} (stat.) \pm 0.036 (syst.),
\end{equation}
which is compatible with the value predicted by the Standard Model within 2.6 standard deviations, and is the most precise measurement of this ratio of branching fractions to date. Further data from an upgrade of the LHCb and from Belle-II are expected within the next years.

Assuming that all radiative corrections are well known, the ratio of the effective FF evaluated with the $\bar{p}p \rightarrow \ell^+\ell^-$ process with $\ell$ = $e, \mu$, could be used to perform a test of the lepton universality at \PANDA at a few percent level:
\begin{equation}
\mathcal{R}_{e\mu} = \frac{|F_p(\bar{p}p \rightarrow \mu^+\mu^-)|}{|F_p(\bar{p}p \rightarrow e^+e^-)|}.
\end{equation}
The estimation of the expected precision of this ratio depends on the expected precision of the effective FF in each of the channels. The studies for the $\bar{p}p \rightarrow \ell^+\ell^-$ reaction were performed at  $q^2$ = 5.4 (GeV/$c$)$^2$ ($p_{beam}$ = 1.7 GeV/$c$) and can be found in Ref. \cite{DmitryAlaaPaper}. The effective FF  is expected to be \cite{Khaneft:2018}
\begin{equation}
|F_p(\bar{p}p \rightarrow e^+e^-)| = 0.1216 \pm 0.0004 \; \mbox{(stat.)} \;\pm 0.0024 \; \mbox{(syst.)}. 
\end{equation}
From that, the total relative uncertainty is obtained as 
\begin{equation}
\Delta |F_p(\bar{p}p \rightarrow e^+e^-)| / |F_p(\bar{p}p \rightarrow e^+e^-)| \sim 2.02\%. 
\end{equation}
For the muon channel, the effective proton FF value
\begin{equation}
|F_p(\bar{p}p \rightarrow \mu^+\mu^-)| = 0.1210 \pm 0.0009 \; \mbox{(stat.)} \; \pm 0.0024 \;\mbox{(syst.)}
\end{equation}
is obtained, so one gets for the ratio
\begin{equation}
\mathcal{R}_{e\mu} = 0.99 \pm 0.03,
\end{equation}
which corresponds to a relative total uncertainty of  $\sim 3\%$. An even better precision would be expected for the lowest  $q^2$ = 5.1 (GeV/$c$)$^2$  value ($p_{beam}$ = 1.5 GeV/$c$), since the signal cross section has higher values.

From these results, it can be concluded that \PANDA will be able to perform a test of a possible violation of the lepton universality ($e$-$\mu$) with a good precision, provided the QED radiative corrections are precisely known for both channels. This calls for a new set of calculations.

\section{Summary}

 A thorough feasibility study for the measurement of the  time-like proton form factors in the    $\bar p p \to \mu^+ \mu^-$ reaction is performed within the PANDARoot framework at four beam momenta between 1.5 and 3.3 GeV/$c$. A method based on multivariate data classification (Boosted Decision Trees) is used to optimize the separation of the signal from the main background channel $\bar p p \to \pi^+ \pi^-$.  Signal to background ratios between 1:5 and 1:13 (background rejection factor of $\sim 10^{-5}$) are achieved.  A subtraction of the residual background events is performed. Assuming an integrated luminosity of 2 fb$^{-1}$ per beam momentum setting, the  statistical precisions of ${|G_{E}|}$, ${|G_{M}|}$ and $R=|G_{E}|/|G_{M}|$ are determined by fitting the angular distributions of the  $\bar p p \to \mu^+ \mu^-$ signal reaction. The main contributions to the systematic uncertainties are studied, determining the precisions of the integrated cross section of the  $\bar p p \to \mu^+ \mu^-$ reaction and the proton effective FF.  The results of the simulations show that the proton form factors can be measured exploiting the  $\bar p p \to \mu^+ \mu^-$ reaction at \PANDA with good precision:  a total relative uncertainty on the measurement of the proton form factor ratio between $5\%$ at 1.5 GeV/$c$  and $37\%$ at 3.3 GeV/$c$, for instance, is expected.

In Ref.\cite{DmitryAlaaPaper}  the feasibility studies for the   $\bar{p}p \rightarrow e^+e^-$ reaction was presented, which will be also used at \PANDA to study the proton FFs in the time-like region. It has been shown  that a suppression of background pollution from $\bar{p}p \rightarrow \pi^+\pi^-$ to the level of a few percent will be possible, which makes background subtraction unnecessary. A very high expected precision of the FF ratio was obtained in these studies with total values up to 1.3$\%$ (stat.) and 3.3$\%$  at $q^2$ = 5.4 (GeV/$c$)$^2$. Compared to these results, the muon channel provides a limited precision due to the relatively low signal efficiency, uncertainty in background subtraction and consequently the additional statistical fluctuations in the signal angular distribution. The measurement of the muon channel is very challenging due to the overwhelming pion background, therefore the precision of the proton FFs determined in these feasibility studies provides an exciting opportunity for the experiment and  the measurement of this channel will offer a very promising contribution to the rich \PANDA physics program.

\section{Acknowledgements}

We acknowledge financial support from 
the Science and Technology Facilities Council (STFC), British funding agency, Great Britain; 
the Bhabha Atomic Research Centre (BARC) and the Indian Institute of Technology Bombay, India; 
the Bundesministerium f\"ur Bildung und Forschung (BMBF), Germany; 
the Carl-Zeiss-Stiftung 21-0563-2.8/122/1 and 21-0563-2.8/131/1, Mainz, Germany; 
the Center for Advanced Radiation Technology (KVI-CART), Groningen, Netherlands; 
the CNRS/IN2P3 and the Universit\'{e} Paris-Sud, France; 
the Czech Ministry (MEYS) grants LM2015049, CZ.02.1.01/0.0/0.0/16 and 013/0001677, 
the Deutsche Forschungsgemeinschaft (DFG), Germany; 
the Deutscher Akademischer Austauschdienst (DAAD), Germany; 
the Forschungszentrum J\"ulich, Germany; 
the European Union's Horizon 2020 research and innovation program under grant agreement No 824093;
the Gesellschaft f\"ur Schwerionenforschung GmbH (GSI), Darmstadt, Germany; 
the Helmholtz-Gemeinschaft Deutscher Forschungszentren (HGF), Germany; 
the INTAS, European Commission funding; 
the Institute of High Energy Physics (IHEP) and the Chinese Academy of Sciences, Beijing, China; 
the Istituto Nazionale di Fisica Nucleare (INFN), Italy; 
the Ministerio de Educacion y Ciencia (MEC) under grant FPA2006-12120-C03-02; 
the Polish Ministry of Science and Higher Education (MNiSW) grant No. 2593/7, PR UE/2012/2, and the National Science Centre (NCN) DEC-2013/09/N/ST2/02180, Poland; 
the State Atomic Energy Corporation Rosatom, National Research Center Kurchatov Institute, Russia; 
the Schweizerischer Nationalfonds zur F\"orderung der Wissenschaftlichen Forschung (SNF), Swiss; 
the Stefan Meyer Institut f\"ur Subatomare Physik and the \"Osterreichische Akademie der Wissenschaften, Wien, Austria; 
the Swedish Research Council and the Knut and Alice Wallenberg Foundation, Sweden;
the Scientific and Technological Research Council of Turkey (TUBITAK) under grant No. 119F094.

 \bibliographystyle{unsrt}
 \bibliography{include/sample_v3}

\begin{thebibliography}{10}

\bibitem{Hof:1956}
R~Hofstadter.
\newblock {\em Rev. Mod. Phys.}, 28:214--254, 1956.

\bibitem{Ak:1968}
A.~Akhiezer and M.~Rekalo.
\newblock {\em Sov. Phys. Dokl.}, (13):572, 1968.

\bibitem{Akhiezer:1974em}
A.~I. Akhiezer and Mikhail.P. Rekalo.
\newblock {\em Sov. J. Part. Nucl.}, 4:277, 1974.
\newblock [Fiz. Elem. Chast. Atom. Yadra4,662(1973)].

\bibitem{Jo:2000}
M.~K. Jones et~al.
\newblock {\em Phys. Rev. Lett.}, (84):1398, 2000.

\bibitem{Ga:2002}
O.~Gayou et~al.
\newblock {\em Phys. Rev. Lett.}, (88):092301, 2002.

\bibitem{Pun:2005}
V.~Punjabi et~al.
\newblock {\em Phys. Rev.}, C(71):055202, 2005.

\bibitem{Pu:2010}
A.~J.~R. Puckett et~al.
\newblock {\em Phys. Rev. Lett.}, 104(242301), 2010.

\bibitem{Puckett:2017flj}
A.~J.~R. Puckett et~al.
\newblock {\em Phys. Rev. C}, 96(5):055203, 2017.

\bibitem{Rosenbluth:1950yq}
M.N. Rosenbluth.
\newblock {\em Phys. Rev.}, 79:615--619, 1950.

\bibitem{Mohr:2008fa}
Peter~J. Mohr, Barry~N. Taylor, and David~B. Newell.
\newblock {\em Rev. Mod. Phys.}, 80:633--730, 2008.

\bibitem{Pohl:2010}
R.~Pohl et~al.
\newblock {\em Nature}, 466:213, 07 2010.

\bibitem{Bezginov:2019mdi}
N.~Bezginov, T.~Valdez, M.~Horbatsch, A.~Marsman, A.~C. Vutha, and E.~A.
  Hessels.
\newblock {\em Science}, 365(6457):1007--1012, 2019.

\bibitem{Gilman:2017hdr}
R.~Gilman et~al.
\newblock arXiv:1709.09753 [physics.ins-det], 2017.

\bibitem{Castellano:1973}
M.~Castellano et~al.
\newblock {\em Nuovo Cim.}, 14 A(1), 1973.

\bibitem{Andreotti:2003}
M.~Andreotti et~al.
\newblock {\em Phys.Lett.}, B(559):20, 2003.

\bibitem{Ambrogiani:1999}
M.~Ambrogiani et~al.
\newblock {\em Phys.Rev. D}, 60(032002), 1999.

\bibitem{Antonelli:1998fv}
A.~Antonelli et~al.
\newblock {\em Nucl. Phys. B}, 517:3--35, 1998.

\bibitem{Bardin:1994am}
G.~Bardin et~al.
\newblock {\em Nucl.Phys.}, B411:3--32, 1994.

\bibitem{Armstrong:1993}
T.~A. Armstrong et~al.
\newblock {\em Phys.Rev. Lett.}, 70:1212, 1993.

\bibitem{Delcourt:1979}
B.~Delcourt et~al.
\newblock {\em Phys.Lett. B}, B(86):395, 1979.

\bibitem{Bisello:1983}
D.~Bisello et~al.
\newblock {\em Nucl.Phys. B}, B(224):379, 1983.

\bibitem{Bisello:1990}
D.~Bisello et~al.
\newblock {\em Z. Phys. C}, C(48):23, 1990.

\bibitem{Ablikim:2005}
M.~Ablikim et~al.
\newblock {\em Phys.Lett.}, B(630):14.

\bibitem{Ablikim:2015}
M.~Ablikim et~al.
\newblock {\em Phys.Rev.}, D(91), 2015.

\bibitem{Pedlar:2005}
T.K. Pedlar et~al.
\newblock {\em Phys.Rev. Lett.}, 95(261803), 2005.

\bibitem{Akhmetshin:2015ifg}
R.~R. Akhmetshin et~al.
\newblock {\em Phys. Lett. B}, 759:634, 2016.

\bibitem{Lees:2013b}
J.P. Lees et~al.
\newblock {\em Phys.Rev. D}, 87(092005), 2013.

\bibitem{Lees:2013}
J.~P. Lees et~al.
\newblock {\em Phys. Rev. D}, 88(032011), 2013.

\bibitem{Ablikim:2019njl}
M.~Ablikim et~al.
\newblock {\em Phys. Rev.}, D99(9):092002, 2019.

\bibitem{Ablikim:2019eau}
M.~Ablikim et~al.
\newblock {\em Phys. Rev. Lett.}, 124(4):042001, 2020.

\bibitem{TomasiGustafsson:2008gq}
Egle Tomasi-Gustafsson and Michail~P. Rekalo.
\newblock Int. Report DAPNIA-04-01, arXiv:0810.4245 [hep-ph], 2008.

\bibitem{Dbeyssi:2011tv}
A.~Dbeyssi, E.~Tomasi-Gustafsson, G.~I. Gakh, and M.~Konchatnyi.
\newblock {\em Nucl.Phys.}, A(894), 2012.

\bibitem{Dubnickova:1995ns}
A.~Z. Dubnickova, S.~Dubnicka, and M.~P. Rekalo.
\newblock {\em Z. Phys.}, C70:473--482, 1996.

\bibitem{Adamuscin:2006bk}
C.~Adamuscin, E.~A. Kuraev, E.~Tomasi-Gustafsson, and F.~E. Maas.
\newblock {\em Phys. Rev.}, C75:045205, 2007.

\bibitem{Guttmann:2013}
J.~Guttmann and M.~Vanderhaeghen.
\newblock {\em Phys. Lett.}, B(719):136--142, 2013.

\bibitem{Sudol:2010}
M.~Sudol et~al.
\newblock {\em Eur. Phys. J. A}, 44:373, 2010.

\bibitem{DmitryAlaaPaper}
W.~Erni B.~Singh et~al.
\newblock {\em Eur. Phys. J. A}, A(52):325, 2016.

\bibitem{Lutz:2009ff}
M.~Lutz et~al.
\newblock \textit{PANDA Physics Performance Report}.
\newblock arXiv:0903.3905v1 [hep-ex], 2009.

\bibitem{Singh:2015}
B.~P. Singh et~al.
\newblock {\em Eur. Phys. J. A}, 51(8):107, Aug 2015.

\bibitem{Singh:2016qjg}
B.~Singh et~al.
\newblock {\em Phys. Rev.}, D95(3):032003, 2017.

\bibitem{FAIR:2006}
{FAIR Baseline Technical Report}.
\newblock 2006.

\bibitem{PandaTDRTarget:2014}
W.~Erni, I.~Keshelashvili, B.~Krusche, et~al.
\newblock \textit{Technical Design Report for the PANDA Internal Targets: The
  Cluster-Jet Target and Developments for the Pellet Target}.
\newblock Technical report, 2014.

\bibitem{PandaTDRMagnet:2009}
W.~Erni, I.~Keshelashvili, B.~Krusche, et~al.
\newblock \textit{Technical Design Report for the PANDA Solenoid and Dipole
  Spectrometer Magnets}.
\newblock Technical report, 2009.

\bibitem{PandaTDRMVD:2012}
W.~Erni, I.~Keshelashvili, B.~Krusche, et~al.
\newblock \textit{Technical Design Report for the PANDA Micro Vertex Detector}.
\newblock Technical report, 2012.
\newblock arXiv:physics.ins-det/1207.6581.

\bibitem{PandaTDRStt:2013}
W.~Erni, I.~Keshelashvili, B.~Krusche, et~al.
\newblock \textit{Technical Design Report for the PANDA Straw Tube Tracker}.
\newblock {\em Eur. Phys. J.}, A(49), 2009.

\bibitem{PandaDIRC:2005}
W.~Erni et~al.
\newblock \textit{Technical Design Report for the PANDA Barrel DIRC Detector}.
\newblock Technical report, 2005.
\newblock arXiv:1710.00684 [physics.ins-det].

\bibitem{PandaTDREMC:2008}
W.~Erni, I.~Keshelashvili, B.~Krusche, et~al.
\newblock \textit{Technical Design Report for the PANDA electromagnetic
  calorimeter}.
\newblock arXiv:physics.ins-det/0810.1216, 2008.

\bibitem{PandaTDRMDT:2012}
W.~Erni, I.~Keshelashvili, B.~Krusche, et~al.
\newblock \textit{Technical Design Report for the PANDA Muon System}.
\newblock Technical report, 2012.

\bibitem{PandaTDRFSC:2016}
W.~Erni, I.~Keshelashvili, B.~Krusche, et~al.
\newblock \textit{Technical Design Report for the PANDA Forward Spectrometer
  Calorimeter}.
\newblock Technical report, 2016.

\bibitem{Zichichi:1962ni}
A.~Zichichi, S.~M. Berman, N.~Cabibbo, et~al.
\newblock {\em Nuovo Cim.}, 24:170--180, 1962.

\bibitem{Brodsky:2007hb}
Stanley~J. Brodsky and Guy~F. de~Teramond.
\newblock {\em Phys. Rev. D}, 77:056007, 2008.

\bibitem{TomasiGustafsson:2001za}
E.~Tomasi-Gustafsson and M.~P. Rekalo.
\newblock {\em Phys.Lett.}, B(504):291--295, 2001.

\bibitem{Shirkov:1997}
D.V. Shirkov et~al.
\newblock {\em Phys.Rev. Lett.}, 79:1209, 1997.

\bibitem{Bian:2015}
A.~Bianconi and E.~Tomasi-Gustafsson.
\newblock {\em Phys. Rev. Lett.}, 114:232301, Jun 2015.

\bibitem{Eisenhandler:1975kx}
E.~Eisenhandler, W.R. Gibson, et~al.
\newblock {\em Nucl.Phys.}, B(96):109, 1975.

\bibitem{Spataro:2012}
S.~Spataro et~al.
\newblock {\em J. Phys.: Conf. Series}, 396:9, 2012.

\bibitem{fairroot}
M.~Al-Turany et~al.
\newblock {\em Journal of Physics: Conference Series}, 396(2):022001, 2012.

\bibitem{Boucher:2011}
J.~Boucher et~al.
\newblock {\em Feasibility studies of the $\bar{p}p \rightarrow \pi^0 e^+e^-$
  electromagnetic channel at PANDA}.
\newblock PhD thesis, 2011.

\bibitem{AlaasThesis}
A.~Dbeyssi.
\newblock {\em \textit{Study of the internal structure of the proton with the
  PANDA experiment at FAIR}}.
\newblock PhD thesis, IPN Orsay, 2013.

\bibitem{Wang:2015ybw}
Ying Wang, Yury~M. Bystritskiy, and Egle Tomasi-Gustafsson.
\newblock {Antiproton-proton annihilation into charged light meson pairs within
  effective meson theory}.
\newblock {\em Phys. Rev.}, C95(4):045202, 2017.

\bibitem{Wang:2017pkc}
Y.~Wang, Yury~M. Bystritskiy, Azad~I. Ahmadov, and Egle Tomasi-Gustafsson.
\newblock {\em Phys. Rev.}, C96(2):025204, 2017.

\bibitem{VandeWiele:2010kz}
J.~Van de~Wiele and S.~Ong.
\newblock {\em Eur. Phys. J.}, A46:291--298, 2010.

\bibitem{Zambrana:2014}
M.~Zambrana et~al.
\newblock Technical report, HIM Mainz, 2014.

\bibitem{Eide:1973tb}
A.~Eide et~al.
\newblock {\em Nucl.Phys.}, B(60):173--220, 1973.

\bibitem{Buran:1976wc}
T.~Buran et~al.
\newblock {\em Nucl.Phys.}, B(116):51, 1976.

\bibitem{Armstrong:1986ng}
T.A. Armstrong et~al.
\newblock {\em Nucl. Phys.}, B284:643, 1987.

\bibitem{White:1994tj}
C.~White, R.~Appel, D.~S. Barton, et~al.
\newblock {\em Phys. Rev.}, D(49):58--78, 1994.

\bibitem{root}
CERN ROOT.
\newblock \textit{https://root.cern.ch/root/html/TROOT.html}.

\bibitem{tmva}
A.~Hoecker et~al.
\newblock \textit{http://tmva.sourceforge.net/}, 09 2013.

\bibitem{Ryd:2005}
A.~Ryd et~al.
\newblock \textit{EvtGen: A Monte Carlo Generator for B-Physics}, 2005.

\bibitem{LumiTechRep}
W.~Erni et~al.
\newblock {\it{Technical Report for the PANDA Luminosity Monitor}}.
\newblock Technical Report.

\bibitem{Pa:2015}
S.~Pacetti, R.~Baldini, and E.~Tomasi-Gustafsson.
\newblock \textit{Proton electromagnetic form factors: Basic notions, present
  achievements and future perspectives.}
\newblock {\em Phys.Rept.}, 550-551:1--103, 2014.

\bibitem{Guttmann:2011b}
J.~Guttmann, N.~Kivel, and M.~Vanderhaeghen.
\newblock {\em Phys. Rev.}, D(83):094021, 2011.

\bibitem{VandeWiele:2013}
J.~van~de Wiele and S.~Ong.
\newblock {\em Eur. Phys. J. A (2013) 49: 18}, A(49):18, 2013.

\bibitem{Gakh:2005}
G.I. Gakh and E.~Tomasi-Gustafsson.
\newblock {\em Nucl.Phys.}, A(761):120--131, 2005.

\bibitem{Akhmetshin:2016}
R.~R. Akhmetshin et~al.
\newblock arXiv:1507.08013v2 [hep-ex], April 2016.

\bibitem{Aaij:2014}
R.~Aaij, B.~Adeva, M.~Adinolfi, et~al.
\newblock {\em Phys. Rev. Lett.}, 113:151601, 2014.

\bibitem{Bo:2007}
C.~Bobeth, G.~Hiller, and G.~Piranishvili.
\newblock {\em JHEP}, 12 2007.

\bibitem{Bouchard:2013}
C.~Bouchard et~al.
\newblock {\em Phys. Rev. Lett.}, 111(162002), 2013.

\bibitem{Bordone:2016}
M.~Bordone, Isidori G., and A.~Pattori.
\newblock {\em Eur. Phys. J.}, C(76):440, 2016.

\bibitem{Khaneft:2018}
A.~Dbeyssi, D.~Khaneft, et~al.
\newblock \textit{Release Note}, 2018.

\end{thebibliography}
%
%
%

\end{document}